\documentclass[pdflatex,sn-nature]{sn-jnl}

\usepackage{graphicx}%
\usepackage{multirow}%
\usepackage{amsmath,amssymb,amsfonts}%
\usepackage{amsthm}%
\usepackage{mathrsfs}%
\usepackage[title]{appendix}%
\usepackage{xcolor}%
\usepackage{textcomp}%
\usepackage{manyfoot}%
\usepackage{booktabs}%
\usepackage{algorithm}%
\usepackage{algorithmicx}%
\usepackage{algpseudocode}%
\usepackage{listings}%

\theoremstyle{thmstyleone}%
\theoremstyle{thmstyletwo}%

\theoremstyle{thmstylethree}%

\makeatletter
\def\@orcid#1{} 
\makeatother

\begin{document}

\title[Article Title]{Duration-modulated neural population dynamics in humans during BMI controls}

\author*[1,2]{\fnm{Fei} \sur{Yin}}\email{fyin@caltech.edu}
\author[1,2]{\fnm{Charles} \sur{Guan}}
\author[1,2]{\fnm{Tyson} \sur{Aflalo}}
\author[1,2]{\fnm{Jorge} \sur{Gamez}}
\author[1,2]{\fnm{Kelsie} \sur{Pejsa}}
\author[3]{\fnm{Emily} \sur{Rosario}}
\author[4,5]{\fnm{Charles} \sur{Liu}}
\author[6]{\fnm{Ausaf} \sur{Bari}}
\author*[1,2]{\fnm{Richard} \sur{Andersen}}\email{richard.andersen@vis.caltech.edu}

\affil*[1]{
    \orgdiv{Division of Biology and Biological Engineering}, 
    \orgname{California Institute of Technology}, 
    \orgaddress{
        \city{Pasadena}, 
        \postcode{91125}, 
        \state{CA}, 
        \country{USA}}}

\affil[2]{
    \orgdiv{T\&C Chen Brain-machine Interface Center}, 
    \orgname{California Institute of Technology}, 
    \orgaddress{
        \city{Pasadena}, 
        \postcode{91125}, 
        \state{CA}, 
        \country{USA}}}

\affil[3]{
    \orgdiv{Research Institute}, 
    \orgname{Casa Colina Hospital and Centers for Healthcare}, 
    \orgaddress{
    \city{Pomona}, 
    \postcode{91767}, 
    \state{CA}, 
    \country{USA}}}

\affil[4]{
    \orgdiv{Department of Neurological Surgery}, 
    \orgname{Keck School of Medicine of USC}, 
    \orgaddress{
        \city{Los Angeles}, 
        \postcode{90033}, 
        \state{CA}, 
        \country{USA}}}

\affil[5]{
    \orgdiv{Neurorestoration Center}, 
    \orgname{Keck School of Medicine of USC}, 
    \orgaddress{
        \city{Los Angeles}, 
        \postcode{90033}, 
        \state{CA}, 
        \country{USA}}}

\affil[6]{
    \orgdiv{Department of Neurosurgery}, 
    \orgname{David Geffen School of Medicine at UCLA}, 
    \orgaddress{
        \city{Los Angeles}, 
        \postcode{90095}, 
        \state{CA}, 
        \country{USA}}}

\abstract{The motor cortex (MC) is often described as an autonomous dynamical system during movement execution. In an autonomous dynamical system, flexible movement generation depends on reconfiguring the initial conditions, which then unwind along known dynamics. An open question is whether these dynamics govern MC activity during brain-machine interface (BMI) control. We investigated MC activity during BMI cursor movements of multiple durations, ranging from hundreds of milliseconds to sustained over seconds. These durations were chosen to cover the range of movement durations necessary to control modern BMIs under varying precision levels. Movements shared their MC initial condition with movements of different durations in the same direction. Long-duration movements sustained MC activity, effectively pausing the neural population dynamics until each movement goal was reached. The difference across durations in MC population dynamics may be attributed to external inputs. Our results highlight the role of sustained inputs to MC during movement.}


\maketitle

\section{Introduction}\label{introduction}
The motor cortex (MC) plays a central role in volitional movements, but its precise mechanism is still debated \citep{Omrani2017-hm}.  MC studies from the 1980s to early 2000s largely focused on representational modeling (RM), identifying movement parameters that correlate directly with neural activity. Studies discovered a wide range of represented parameters, including movement direction, speed, force, muscle activity, and posture \citep{Evarts1968-rw, Georgopoulos1986-sm, Georgopoulos1992-vd, Kakei1999-nk, Moran1999-bh, Paninski2004-sh, Aflalo2006-wr}. However, evidence for a literal parameter representation remains inconclusive. Multiple movement parameters were often represented simultaneously, and these representations shifted across task contexts \citep{Aflalo2007-wm, Scott2008-hb, Omrani2017-hm} or even within a single movement \citep{Churchland2007-ed}, leading some researchers to suggest that this parameter search was misleading \citep{10.1017/cbo9780511529788.008, Scott2008-hb, Omrani2017-hm}.

The autonomous dynamical systems hypothesis (aDSH) emerged as a response to the limitations of representational modeling (RM) in explaining MC activity. The aDSH states that the motor cortex comprises a pattern generator, where preparatory activity sets the initial condition and movement execution activity unfolds along predictable dynamics \citep{Churchland2012-hp, Shenoy2013-dn, Pandarinath2015-im, Pandarinath2018-kr, Pandarinath2018-of, Sussillo2015-cz, Vyas2020-an}. The aDSH has been demonstrated most comprehensively with able-bodied non-human primate subjects performing a reach tasks \citep{Churchland2010-ck, Churchland2012-hp, Kaufman2014-dl, Elsayed2016-lc}. In these tasks, movement preparation initializes the neural state of the dorsal premotor (PMd) and motor cortex, which then determines quasi-oscillatory MC activity during reach execution \citep{Pandarinath2018-kr}. Recurrent neural network models suggest that local recurrence could implement MC’s brief oscillations \citep{Sussillo2015-cz, Pandarinath2018-kr, Russo2018-bb, OShea2022-gu, Saxena2022-tz}, which provide a basis set for generating multiphasic muscle activity \citep{Churchland2012-hp, Sussillo2015-cz}.  

Mathematically, a dynamical system can be described by a function $f$ mapping the system state $x$ and external inputs $u$ to the instantaneous change in state $\frac{dx}{dt}$:

\begin{equation}
    \frac{dx}{dt} = f(x(t), u(t))
\label{eqt:dynamical_system_f}
\end{equation}

where $t$ indicates time. Frequently \citep{Shenoy2013-dn, Sauerbrei2020-sk}, the system state is defined as the MC firing rate vector $\mathbf{r}$, and $f$ is constrained to be an additive linear function $h$ that models local recurrence. The input $\mathbf{u}$ can include external stimuli and the firing rates of other brain areas. In this simplification, Equation \ref{eqt:dynamical_system_f} can be written as: 

\begin{equation}
    \frac{d\mathbf{r}}{dt} = h(\mathbf{r}(t)) + \mathbf{u}(t)
\label{eqt:dynamical_system_hu}
\end{equation}

The aDSH posits that the system (MC) is self-contained during the execution of prepared movements \citep{Kalidindi2021-qz, Kao2015-sq}. That is, $u$ is negligible during movement execution, further simplifying Equation \ref{eqt:dynamical_system_hu}  into: 

\begin{equation}
    \frac{d\mathbf{r}}{dt} = h(\mathbf{r}(t)) = \mathbf{Ar}(t) + \mathbf{b}
\label{eqt:dynamical_system_Ab}
\end{equation}

where $\mathbf{A}$ is the linear dynamics matrix and $\mathbf{b}$ is an offset vector. 

The autonomous dynamical systems hypothesis explained several phenomena that had puzzled the RM framework. For example, neurons appear to change representational tuning with time because they generate a temporal basis rather than directly driving the output \citep{Shenoy2013-dn, Pandarinath2018-of}. And apparent modulation by multiple movement parameters was not a literal parameter representation but rather a basis set for downstream muscle read-out.  

Three related claims follow from the autonomous dynamical systems hypothesis. First, because the local MC dynamics $h$ are stable \citep{Gallego2020-ai} and external inputs $\mathbf{u}$ are hypothesized to be negligible, generating different neural trajectories $\mathbf{r}(t)$ is solely determined by specifying different initial conditions $\mathbf{r}(t=0)$ \citep{Churchland2012-hp, Pandarinath2018-kr, Vyas2020-an}. Second, these dynamics cause movement; specifically, neural perturbations disrupt movement if and only if the perturbation alters the task dynamics subspace \citep{OShea2022-gu}. Third, under the aDSH, MC dynamics are agnostic to external input, including sensory feedback. In the absence of that feedback, the circuit cannot correct errors once a movement has begun. Since motor circuits are not noise-free systems, accurate execution depends on the intrinsic MC dynamics being robust to noise. Noise-robust autonomous dynamical systems exhibit low tangling (described in MC by \cite{Russo2018-bb}) (i.e., smooth flow fields) and low divergence (described in the supplementary motor area, but notably not in MC, by \cite{Russo2020-wb}).  

A key open question is whether the same autonomous dynamical principles govern MC activity during brain-machine interface (BMI) control. Movement BMIs aim to restore movement to people with motor disabilities by decoding motor intent directly from neural activity \citep{Hochberg2006-qn, Hochberg2012-ru, Collinger2013-ws, Aflalo2015-ik, Gilja2015-kg, Wodlinger2015-kb, Bouton2016-at, Ajiboye2017-un, Robinson2025-jx}. BMI cursor control has long been considered analogous to able-bodied reaching \citep{Serruya2002-fj, Taylor2002-ry, Hochberg2006-qn, Hwang2013-lu, Golub2016-xg, Inoue2018-de}. Given the success of aDSH in explaining MC activity during able-bodied reaching, this analogy suggests that aDSH-based modeling could improve human BMI control of assistive devices \citep{Kao2015-sq, Pandarinath2018-of}. Notably, beyond planar reaching, aDSH-based models have already been extended to grasping, locomotion, and even speech-related movements \citep{Hall2014-gp, Stavisky2019-hg, Kalidindi2021-qz}, reinforcing their potential as a general framework for diverse BMI applications.

However, the aDSH has also faced critiques, including alternative interpretations of the observed population structure \citep{Lebedev2019-qs, Kalidindi2021-qz} and evidence that its principles may not generalize to more complex or interrupted movements such as dexterous actions \citep{Sauerbrei2020-sk, Suresh2020-qp} or stopping behaviors \citep{Russo2020-wb}. From the perspective of applying aDSH to BMI control, a core limitation is the assumption that inputs like sensory feedback are negligible. Compared to the able-bodied movements usually modeled by aDSH, BMI movements are controlled by a small fraction of neurons in MC, so BMI control is slower, less precise \citep{Shanechi2016-xi, Willett2017-bn}, and prone to nonstationarities \citep{Jarosiewicz2015-fs}. As a result, high-performance BMI control benefits from rapid real-time feedback \citep{Gilja2012-lk, Shanechi2016-xi, Shanechi2017-ms, Willett2018-oc, Willett2019-op}. During able-bodied movement, the motor cortex reflects proprioceptive signals that are missing during BMI control \citep{Stavisky2018-yk}. Taken together, behavioral dynamics and sensory inputs differ substantially between BMI control and able-bodied arm reaching. 

Dynamical systems modeling promises to improve neural decoding performance \citep{Kao2015-sq, Pandarinath2018-of, Gallego2020-ai, Karpowicz2022-fu}. Given the difference in behavioral dynamics between able-bodied reaching and BMI control, an open question is how well the aDSH applies to BMI movements across different applications. To examine whether motor cortex (MC) activity during BMI control can be explained by autonomous dynamics alone or also reflects external drive, we designed variants of a standard cursor task that separately emphasized ballistic and sustained control. When the BMI participant attempted ballistic reaches analogous to previous aDSH studies, MC activity reproduced previously described rotational dynamics. When the BMI participant sustained cursor movement for longer durations, the initial conditions were similar, yet the neural trajectories paused during the sustained period of movements. This divergence indicates that additional inputs beyond local MC contribute to sustained activity, inconsistent with the autonomous dynamical systems hypothesis. These inputs to MC were present in the absence of somatosensory and proprioceptive feedback, indicating that MC receives inputs from other brain areas during relatively simple behaviors. Although autonomous dynamics may exist during able-bodied ballistic movements, MC activity during sustained BMI movements is not strictly constrained by autonomous dynamics.

\section{Results}\label{results}
\subsection{Intracortical brain-machine interface (BMI) cursor control}
\begin{figure}[ht!]
\centering
\includegraphics[width=\linewidth]{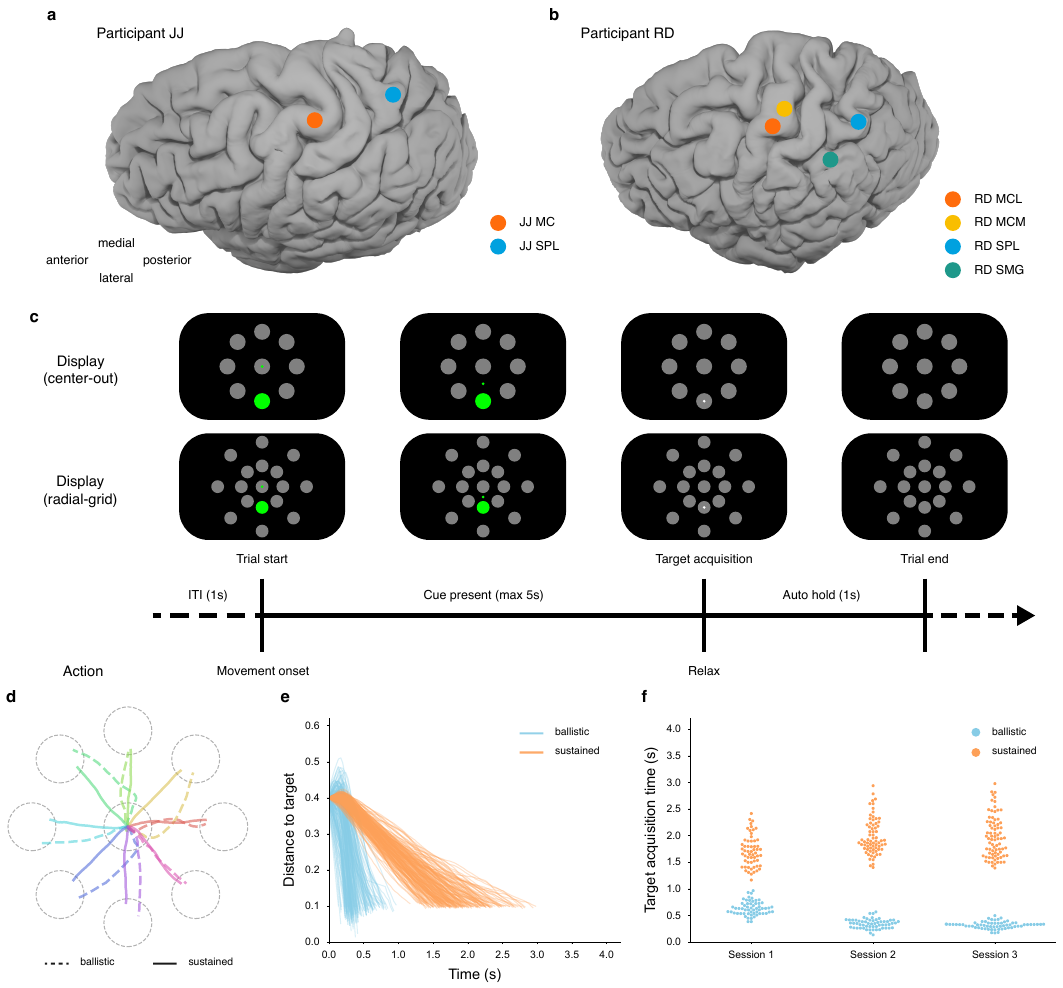}
\caption{Locations of implanted arrays, trial phases, and behavioral data. \textbf{a}, Implant locations in Participant JJ. He was implanted with two 96-channel NeuroPort Utah electrode arrays (Blackrock Microsystems) in the left motor cortex (MC) near the hand knob area and the superior parietal lobule (SPL) of the left posterior parietal cortex (PPC). \textbf{b}, Implant locations of Participant RD. He was implanted with four 64-channel NeuroPort Utah electrode arrays in the hand knob area of the left MC as well as the SPL and supramarginal gyrus (SMG) of the left PPC. \textbf{c}, Trial phases of the (ballistic vs sustained) center-out and (near vs far) radial-grid tasks. The participant sees a screen with targets radially arranged around a center reference. As a trial starts, one of the targets is cued (turns green), instructing the participant to move the cursor toward that target. When the cursor makes contact with the cued target, the participant relaxes, and the cursor is then automatically gravitated toward the center of that target. Consecutive trials are separated by a 1 s inter-trial interval (ITI). \textbf{d}, Example single-trial cursor trajectories during ballistic and sustained conditions of the center-out task. The trajectories are color-coded by target. Targets are shown as circles with dashed lines. \textbf{e}, Distance to target over time for Participant JJ's center-out task trials. Each trace represents a trial. \textbf{f}, Target acquisition times for Participant JJ's center-out task trials. Each marker represents a trial.}
\label{fig1:brains tasks & behavioral data}
\end{figure}

We recorded neural activity from microelectrode arrays implanted in participants JJ's and RD's hand knob of the left motor cortex (MC) and posterior parietal cortex (PPC) (Figure \ref{fig1:brains tasks & behavioral data}a,b) while they completed various brain-machine interface (BMI) 2-D cursor tasks. To calibrate a BMI decoder, the participants observed a computer-controlled cursor perform the center-out-and-back task with 8 targets. Simultaneously, participants JJ and RD attempted to move their right thumb and right index finger, respectively, as if they were controlling the cursor via an analog game controller stick. JJ and RD lost control of their hands after cervical spinal cord injuries, but they attempted to move their effectors as though they were not paralyzed. Using data from this attempted movement calibration task, we trained a decoder to map the predicted cursor velocity from threshold crossing neural firing rates. Next, the participants repeated the center-out-and-back task under closed-loop BMI control. After each block, we reduced the level of computer assistance (see Methods) and updated the decoder. They employed the same attempted movement strategy during online BMI control; we use the terms “BMI control” and “movement” interchangeably in this context. Although all arrays were used for online BMI control, our focus was to better understand the autonomous dynamical systems hypothesis (aDSH), which describes the motor cortex (MC). Therefore, we only analyzed MC recordings.

\subsection{Ballistic and sustained BMI movements}

The autonomous dynamical systems hypothesis (aDSH) is typically studied in non-human primates performing ballistic arm reaches, with movement durations on the order of 400 ms \citep{Churchland2012-hp, Pandarinath2018-kr}, but comparably precise BMI movements can take several seconds \citep{Willett2017-bn}. BMI-useful models of MC activity should apply to movements of both ballistic and sustained timescales. To replicate ballistic and sustained reaches, the participants performed the center-out BMI task (Figure \ref{fig1:brains tasks & behavioral data}c) under two decoder gain parameters. For ballistic reaches, the decoder gain was set so that movements completed within roughly 500 ms. For sustained reaches, we decreased the decoder gain to require sustained movement of more than 1.5 seconds. We wanted to investigate reaches analogous to well-practiced, able-bodied reaching, but BMI control is imprecise at the high decoder gains \citep{Willett2017-bn} used in the ballistic condition. Therefore, we partially assisted ballistic cursor control (see Methods); the assistance level was kept small to ensure that cursor movement was still primarily driven by neural activity. Furthermore, participants were not required to hold the cursor at the target. They were instructed to relax when they reached the target. Ballistic and sustained reaches were grouped into respective trial blocks. Trial blocks were interleaved. 

Under both gain parameters, the participants acquired the targets consistently (Figure \ref{fig1:brains tasks & behavioral data}d-f; Video 1). As designed, ballistic (BMI) movement trials took a roughly similar amount of time as previous non-human primate studies \citep{Churchland2007-ed, Churchland2012-hp} (acquisition time, Participant JJ sessions: 0.43 s ± SD 0.17 s; Participant RD sessions: 0.79 s ± SD 0.16 s). As designed, sustained movement trials were substantially longer (acquisition time, Participant JJ sessions: 1.91 s ± SD 0.37 s; Participant RD sessions: 2.07 s ± SD 0.41 s; 95\% confidence interval of mean difference across conditions (ballistic vs sustained): [1.42, 1.53] s for Participant JJ sessions, [1.24, 1.32] s for Participant RD sessions). These patterns were consistent across sessions (Figure \ref{fig:supp 1 - behavioral data}a,b).

\begin{figure}[h]
\centering
\includegraphics[width=\linewidth]{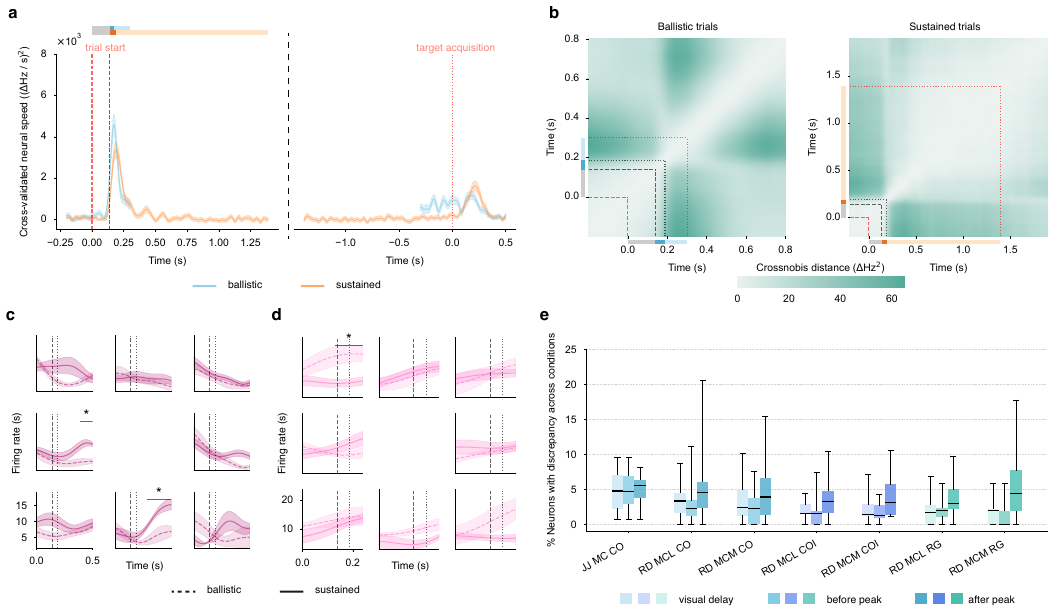}
\caption{Multiphasic neural dynamics during BMI movements and cross-condition discrepancy in single neurons. \textbf{a}, Cross-validated neural speed during ballistic and sustained trials in Participant JJ's center-out sessions. Trials of each condition are truncated to the shortest trial duration observed for that condition. The error bands show 95\% confidence intervals across trials. Left: Trials are aligned to the start time marked by the red dashed line. The black dashed line (left) marks the peak onset time (see Methods). The two horizontal bars above mark the visual delay (gray), transient (dark blue and orange), and steady (light blue and orange) phases. Data includes 200 ms before trial start. Right: Trials are aligned to the target acquisition time marked by the red dotted line. Data includes 500 ms after target acquisition. \textbf{b}, Crossnobis distance across time for ballistic (left) and sustained (right) trials in Participant JJ's center-out sessions. Data includes 200 ms before trial start and 500 ms after target acquisition. Trials of each condition are truncated to the shortest trial duration observed for that condition. The red dashed lines mark the trial start time. The red dotted lines mark the target acquisition time of the shortest trial of each condition. The black dashed lines and dotted lines mark the peak onset times and peak times (see Methods), respectively, in \textbf{a} of each condition. The horizontal and vertical bars beside axis ticks mark the visual delay (gray), transient (dark blue and orange), and steady (light blue and orange) phases. \textbf{c,d}, Example single neuron responses to each target in Participant JJ's center-out sessions. The error bands show SEMs across trials. The dashed and dotted vertical lines mark the peak onset times and peak times, respectively. Significant differences between ballistic and sustained trials are marked (Mann–Whitney U test with false discovery rate correction, *$P < 0.05$). \textbf{e}, Percent of neurons with significant cross-condition discrepancy across targets in all sessions from both participants. Blues, indigos, and greens denote the gain-scaling, interleaved gain-scaling, and distance-scaling tasks, respectively. Lighter to darker shades of each color correspond to the visual delay (see Methods), before-peak, and after-peak windows, respectively.}  
\label{fig2:single and population neurons}
\end{figure}

\subsection{Comparing ballistic and sustained BMI movements reveals multiphasic dynamics}

We computed the cross-validated neural speed (see Methods) for MC activity during ballistic and sustained BMI movements (Figure \ref{fig2:single and population neurons}a; Figure \ref{fig:supp 2 - neural speeds and peak statistics}a,b). Cross-validated neural speed rose sharply during movement onset (Participant JJ: mean $\Delta = 5.26$ kHz$^2$s$^{-1}$, $P < 0.0001$, Cohen’s $d = 2.10$; Participant RD: mean $\Delta = 4.53$ kHz$^2$s$^{-1}$, $P < 0.0001$, Cohen’s $d = 1.73$; Wilcoxon signed-rank test) but then dropped by about 93\% and stabilized near ITI baseline (Participant JJ: mean $\Delta = 0.47$ kHz$^2$s$^{-1}$, $P < 0.0001$, Cohen’s $d = 0.39$; Participant RD: mean $\Delta = 0.27$ kHz$^2$s$^{-1}$, $P < 0.0001$, Cohen’s $d = 0.24$; Wilcoxon signed-rank test), even as the movements continued. This indicates that the neural activity maintained a constant, stationary representation of intent while sustaining the BMI movement. We will refer to the movement onset with high neural speed as the transient phase and the subsequent movement continuation with near-baseline neural speed as the steady phase. Ballistic trials completed earlier than sustained trials, resulting in a less prominent steady phase (Participant JJ: mean durations 0.316 s vs 1.833 s, $\Delta = -1.517$ s, 95 \% CI $-1.548$ to $-1.486$ s, $P < 0.0001$, Cohen’s $d = -5.01$; Participant RD: 0.494 s vs 1.719 s, $\Delta = -1.225$ s, 95 \% CI $-1.244$ to $-1.207$ s, $P < 0.0001$, Cohen’s $d = -3.54$; Mann–Whitney U tests). The initial peak in neural speed occurred at similar times in ballistic and sustained trials (Participant JJ: $\Delta = -0.016$ s, 95\% CI $-0.026$ to $-0.007$ s, $P < 0.0001$, Cohen's $d = -0.39$; Participant RD: $\Delta = -0.022$ s, 95\% CI $-0.033$ to $-0.011$ s, $P = 0.00305$, Cohen's $d = -0.20$; Mann-Whitney U tests) (Figure \ref{fig:supp 2 - neural speeds and peak statistics}m,n). Comparing the two participants, the initial peak occurred about 170 ms earlier ($\Delta = -0.170 s$, 95\% CI $-0.177$ to $-0.163$ s; $P < 0.0001$, Mann-Whitney U test; Cohen's $d = -1.77$) in Participant JJ than in Participant RD under both conditions (Figure \ref{fig:supp 2 - neural speeds and peak statistics}m,n). This is likely due to the difference in reaction time between the two participants.

The crossnobis distance (see Methods) of MC activity between all pairs of time indices during BMI movements (Figure \ref{fig2:single and population neurons}b; Figure \ref{fig:supp 2 - neural speeds and peak statistics}g,h) paints a more holistic picture. In particular, the crossnobis distance between the pre-movement period (ITI and visual delay period) and the steady phase is large, indicating that they do not share tuning profiles. Furthermore, the crossnobis distance is relatively smaller within the steady phase. In other words, it is not just that the participants were inactive during the movement or that coding is purely phasic. Instead, their neural activities were actively sustained during that time. Taken together, when a BMI user attempts to move and sustain motor intent, MC activity appears to consist of an initial dynamic phase and a subsequent steady phase until the target is reached.

\subsection{Mid-movement cross-condition discrepancies in single-neuron and population activity suggest input-driven dynamics}

Task condition modulated the neural response. The aDSH attributes variation in movement to differences in the pre-movement neural state. However, if the onset of such discrepancies takes place after movements have begun, it would suggest that inputs to MC are no longer negligible, contradicting the aDSH. We found neurons that exhibit significant discrepancies in tunings to the same target under different conditions during movements (Figure \ref{fig2:single and population neurons}c,d). We will refer to this kind of discrepancies as mid-movement cross-condition discrepancies. In particular, across participants and tasks, more neurons exhibited mid-movement cross-condition discrepancies during the steady phase than the visual delay phase or the transient phase (Figure \ref{fig2:single and population neurons}e; Figure \ref{fig:supp 3 - perc neurons discrepancy}). This suggests that task-relevant inputs were being integrated during the steady phase.

\begin{figure}[h]
\centering
\includegraphics[width=0.99\linewidth]{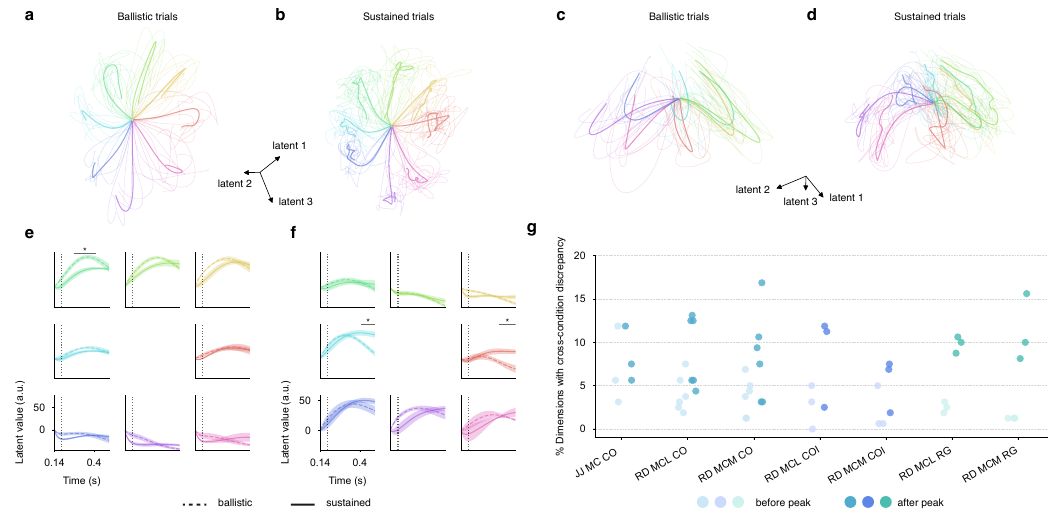}
\caption{Cross-condition discrepancy in population neurons. \textbf{a,b}, Latent neural trajectories from fitting a 3-dimensional linear dynamical system (LDS) to ballistic (\textbf{a}) and sustained (\textbf{b}) trials jointly. Colors represent different targets. Thin trajectories show single trials. Thick trajectories show target averages after time-resampling. \textbf{c,d}, The same LDS latent neural trajectories as in \textbf{a} and \textbf{b} but from a different viewing angle. \textbf{e,f}, Example dimensions of latent neural trajectories separated by target. The error bands show SEMs across trials. Vertical dotted lines mark peak times (see Methods). Significant differences between ballistic and sustained trials are marked (Mann–Whitney U test with false discovery rate correction, *$P < 0.05$). \textbf{g}, Percentage of target dimensions from a 20-dimensional LDS with significant cross-condition discrepancy. Blues, indigos, and greens denote the three tasks. Lighter to darker shades of each color represent the before-peak and after-peak windows, respectively. Each dot represents an experimental session.}
\label{fig3:LDS and population neuron discrepancy}
\end{figure}

Similar patterns are present on the population level. We fit a linear dynamical system (LDS), a proxy implementation of the aDSH, to ballistic and sustained trials jointly. LDS models the data as:

\begin{align}
    \mathbf{x}_{t+1} &= \mathbf{Ax}_t + \mathbf{b} + \mathbf{v}_t  \\
    \mathbf{y}_{t} &= \mathbf{Cx}_t + \mathbf{d} + \mathbf{w}_t
    \label{eqt:LDS}
\end{align}

where $\mathbf{A}$ and $\mathbf{b}$ are the dynamics matrix and bias, $\mathbf{C}$ and $\mathbf{d}$ are the emission matrix and bias, $\mathbf{v}_t$ and $\mathbf{w}_t$ are zero mean Gaussian noise terms, $\mathbf{y}_{t}$ is the observation (MC firing rates), and $\mathbf{x}_t$ is the latent state. When we train an LDS with some high-dimensional sequential data, we effectively reduce the dimensionality to the number of latent states. Note that $\mathbf{A}$ and $\mathbf{b}$ do not change with time.

LDS produced latent trajectories that encode the radial structure of the center-out task (Figure \ref{fig3:LDS and population neuron discrepancy}a-d). Comparing ballistic and sustained trajectories per dimension and target, mid-movement cross-condition discrepancies are observed (Figure \ref{fig3:LDS and population neuron discrepancy}e,f). The percentage of target dimensions exhibiting cross-condition discrepancies increases from the transient phase (before peak) to the steady phase (after peak) to varying extents across participants, sessions, and brain areas (Figure \ref{fig3:LDS and population neuron discrepancy}g).

\begin{figure}[h]
\centering
\includegraphics[width=0.99\linewidth]{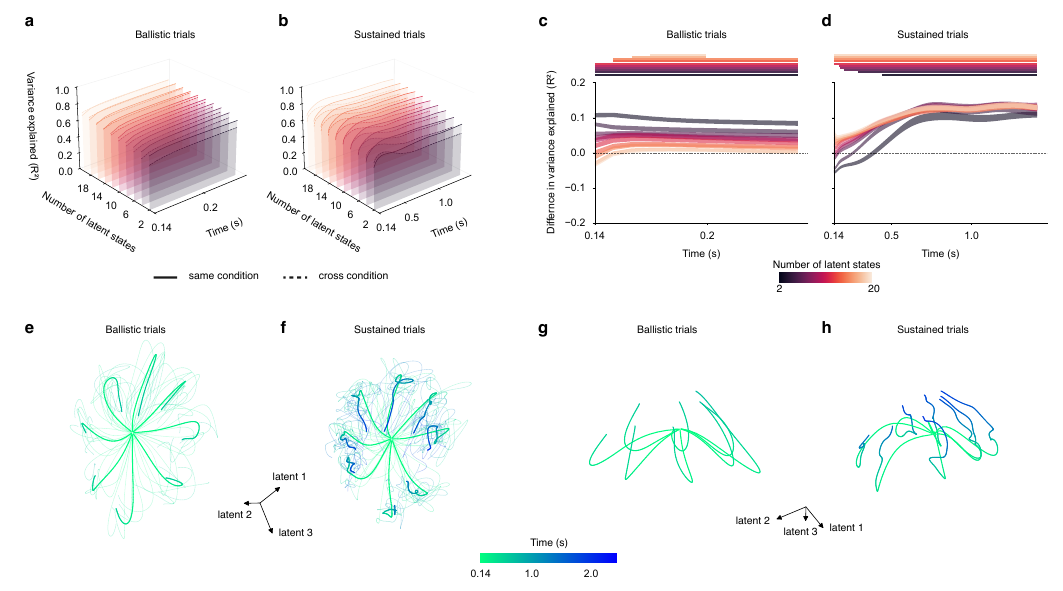}
\caption{Task condition modulated the neural response in LDS latent spaces, showing additional evidence of inputs. \textbf{a,b}, Cross-validated inference performance of LDS models trained on ballistic (\textbf{a}) and sustained (\textbf{b}) trials. Solid and dashed traces denote same- and cross-condition inference, respectively. (e.g., Cross-condition inference of ballistic trials means using a model trained on sustained trials to infer ballistic trials.) \textbf{c,d}, SEMs of difference in performance between same-condition and cross-condition inference of ballistic (\textbf{c}) and sustained (\textbf{d}) trials. Positive values on the y-axis denote better performance achieved by same-condition inference. Horizontal bars indicate significantly better performance achieved by same-condition inference (Wilcoxon signed-rank test with false discovery rate correction, $P < 0.05$). \textbf{e,f}, Latent neural trajectories color-coded by time since the start of trials. Thin trajectories show single trials. Thick trajectories show target averages after time-resampling. \textbf{g,h}, Same as \textbf{e} and \textbf{f} but from a different viewing angle and without single trials.}
\label{fig4:LDS analyses}
\end{figure}

When we compare same‑condition with cross‑condition LDS inference across latent dimensionalities and time, same‑condition models consistently perform better (Figure \ref{fig4:LDS analyses}a-d; Figure \ref{fig:supp 4 - LDS cross speed inference}a,b). However, as the number of latent states increases, and the model becomes more complex, the advantage of same-condition inference diminishes in ballistic trials but not in sustained trials (their steady phase in particular). Because the models predict the transient phase well and fail after transitioning to the steady phase, the error must arise from factors that change once the movement is already underway. A reasonable speculation is that new inputs — visual, cognitive, or error-related — are integrated into MC during the steady phase, steering the neural trajectory away from the purely autonomous path assumed by LDS and the aDSH. 

Time-coloring the LDS trajectories shows that ballistic and sustained movements begin on comparable regions of a low‑dimensional manifold, yet diverge in their temporal evolution: ballistic paths skim the outer “ring” (steady phase) and terminate quickly, whereas sustained paths linger on the same ring for majority of the movement (Figure \ref{fig4:LDS analyses}e-h; Figure \ref{fig:supp 5 - LDS time gradient trajectories}a-n). Such unequal dwell times, coupled with the spatial discrepancies quantified earlier (Figure \ref{fig3:LDS and population neuron discrepancy}e-g), could in principle arise from (i) condition‑specific changes in the local gradient of the flow field, or (ii) condition‑specific external inputs that modulate traversal speed along a shared field.

The linear LDS offers only a single, time‑invariant flow field, so it cannot simultaneously accommodate rapid time‑warping and condition‑specific curvature; this limitation makes (ii) the parsimonious explanation, which contradicts the aDSH. Nonetheless, to rule out the possibility that a non‑linear autonomous system could account for the data without inputs, we next fit a recurrent switching linear dynamical system (rSLDS) that allows piece‑wise linear dynamics while still remaining autonomous.

\subsection{Modeling multiphasic dynamics with switching linear dynamical systems}

\begin{figure}
\includegraphics[width=0.99\linewidth]{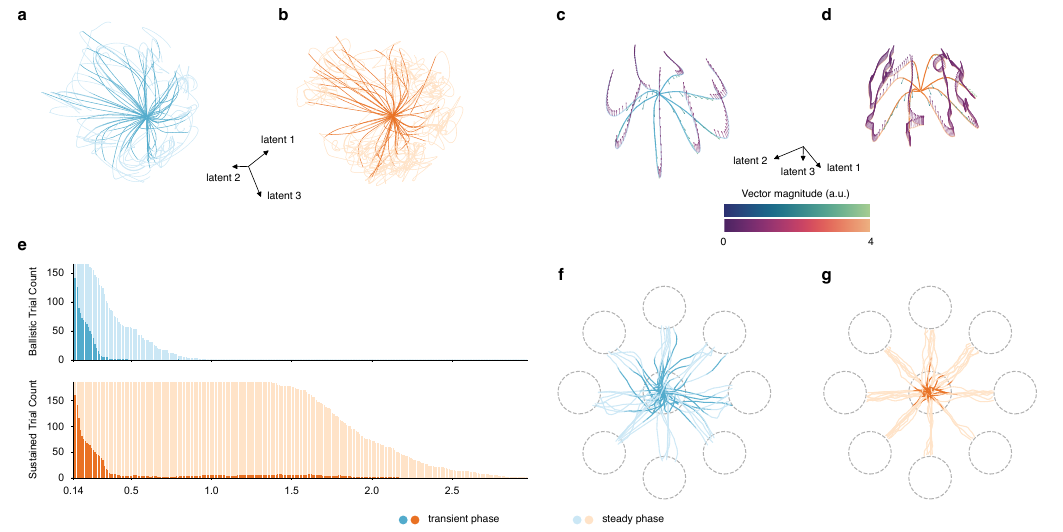}
\caption{Time and discrete state visualization in rSLDS (C3D2) results. \textbf{a,b}, Latent neural trajectories color-coded by discrete states. Each trajectory represents a trial. \textbf{c,d}, Same as \textbf{a} and \textbf{b} but from a different viewing angle and without single trials. In addition, flow field vectors learned by rSLDS are overlaid on the trajectories. Transient phase and steady phase flow field vectors are represented by cold and warm color maps, respectively. \textbf{e}, Trial counts of rSLDS (C3D2) discrete states over time grouped by task condition. \textbf{f,g}, Behavioral cursor trajectories during ballistic \textbf{(f)} and sustained \textbf{(g)} trials overlaied with discrete states. Each trajectory represents a trial. Circles with dashed lines represent the targets.}
\label{fig5:rSLDS analyses}
\end{figure}

Motor cortex activity is clearly multiphasic: a high‑acceleration transient phase is followed by a slower steady phase. When fitted with an LDS, the steady‑phase differs between ballistic and sustained trials in both latent trajectories and time parameterizations (Figures \ref{fig3:LDS and population neuron discrepancy}e-g, Figure \ref{fig4:LDS analyses}).  
One interpretation is that the underlying flow field is nonlinear but still autonomous; another is that a common flow field is modulated by condition‑specific inputs, thereby contradicting the aDSH. 

Recurrent switching linear dynamical systems (rSLDS) \citep{pmlr-v54-linderman17a} provide a principled way to adjudicate between these possibilities: by stitching together piece‑wise linear segments, they approximate a broad family of smooth nonlinear autonomous systems without introducing explicit input terms or condition labels. If nonlinearity alone explains the condition-dependent differences, an rSLDS should require three or more discrete states: one for the transient phase and at least two distinct steady‑phase (for ballistic and sustained trials, respectively) to fit the data optimally. Conversely, if a two-state model suffices, residual condition-linked errors must be captured by external inputs.

We therefore trained rSLDS models on the same MC firing rates, sweeping the number of continuous states (C) and discrete states (D). An rSLDS is defined as:

\begin{align}
    \mathbf{x}_{t+1} &= \mathbf{A}_{z_t} \mathbf{x}_t + \mathbf{b}_{z_t} + \mathbf{v}_t \\
    \mathbf{y}_{t} &= \mathbf{C}_{z_t} \mathbf{x}_t + \mathbf{d}_{z_t} + \mathbf{w}_t
    \label{eqt:SLDS}
\end{align}

Similar to an LDS, $\mathbf{A}_{z_t}$ and $\mathbf{b}_{z_t}$ are the dynamics matrix and bias, $\mathbf{C}_{z_t}$ and $\mathbf{d}_{z_t}$ are the emission matrix and bias, $\mathbf{v}_t$ and $\mathbf{w}_t$ are zero mean Gaussian noise terms, $\mathbf{y}_t$ is the observation (MC firing rates), and $\mathbf{x}_t$ is the latent state. However, with the addition of $z_t$, the discrete state that varies with time, an rSLDS allows the latent dynamics to switch between $z$ sets of dynamics and emission parameters at data-driven switch times. When $z$ equals 1, rSLDS becomes equivalent to an LDS. For clarity, we will refer to ``latent state" as ``continuous state" to distinguish it from ``discrete state" hereafter. 

Fitting an rSLDS with three continuous states and two discrete states (C3D2) yields trajectories (Figure \ref{fig5:rSLDS analyses}a-d) that closely mirror the LDS solution (Figure \ref{fig3:LDS and population neuron discrepancy}a-d). When colored by discrete state, the turning of each rSLDS trajectory aligns with the transient-to-steady transition identified earlier (Figure \ref{fig5:rSLDS analyses}a,b). Overlaying target-averaged trajectories with the learned flow-field vectors reveals that steady-phase vectors have smaller magnitudes than transient-phase vectors under both conditions. This indicates that the population enters a slow-moving attractor during sustained BMI movements, reinforcing the idea that motor cortex dynamics decelerate once the movement enters its steady phase. However, steady-phase data points cluster more tightly in sustained trials than in ballistic trials (Figure \ref{fig5:rSLDS analyses}c,d; Figure \ref{fig:supp 6 - rSLDS discrete state trajectories}a-n), yet their flow field vectors have similar magnitudes. This suggests that inputs have a significant influence on the slow-moving attractor corresponding to the steady phase in the C3D2 rSLDS. 

The fraction of trials in each discrete state over time (Figure \ref{fig5:rSLDS analyses}e; Figure \ref{fig:supp 7 - rSLDS discrete state over time}) shows a consistent pattern: most trials under both conditions begin in the transient phase, switch after similar durations, and then remain in the steady phase. Because sustained trials are longer, the steady phase dominates their time course. Overlaying discrete states on individual cursor traces (Figure \ref{fig5:rSLDS analyses}f,g) shows how the latent state maps onto each trial: the transient phase spans the high-acceleration segment of the reach, while the steady phase corresponds to low-acceleration hold and correction. The transient-state durations are similar in ballistic and sustained trials (Participant JJ: mean $\Delta = -0.0102$ s, $P = 0.297$, Cohen’s $d = -0.16$; Participant RD MCL: mean $\Delta = 0.0222$ s, $P < 0.0001$, Cohen’s $d = 0.28$; Participant RD MCM: mean $\Delta = -0.0056$ s, $P = 0.362$, Cohen’s $d = -0.11$; Mann-Whitney U test), meaning movements under different gains undergo similar accelerations.

Could additional nonlinearity improve the model's fit? To find out, we varied the number of discrete states while sweeping the continuous‑state dimension. To evaluate the models' performance, we used four metrics: position error decoding \citep{Willett2018-oc}, dynamical state update process (DSUP) ratio \citep{Kao2015-sq}, forecast inference, and dimensional entropy difference (see Methods)(Figure \ref{fig:supp 8 - numerical results}). The decoding and inference performance are largely influenced by the number of continuous states, rather than the number of discrete states (nonlinearity) (Figure \ref{fig:supp 8 - numerical results}a,c,e,g). Increasing the nonlinearity helps the model explain more dynamics in the data represented by DSUP ratio (Figure \ref{fig:supp 8 - numerical results}b,f). However, the resulting trial-level distributions of discrete states fail a critical consistency test: with $D\!\ge\!3$, the discrete states were assigned sporadically and lack a stable time‑course across trials as quantified by the dimensional entropy difference (Figure \ref{fig:supp 8 - numerical results}d,h). Thus, additional nonlinear segments do not carve out separate gain‑specific attractors. Therefore, the residual ballistic–sustained discrepancies likely arise from condition‑specific inputs that modulate traversal speed and curvature within the shared two‑state flow.

\subsection{Interleaved gain-scaling and distance-scaling tasks produced similar dynamics}

The main task was block-interleaved, allowing participants to anticipate the cursor's gain for the next movement before trial onset. This may lead to different behaviors and neural dynamics. To control for the effect of anticipation, we experimented with an interleaved version of the ballistic and sustained center-out while leaving the gain parameters unchanged. The participant acquired the targets consistently (ballistic trial acquisition time: 0.84 s ± SD 0.20 s; sustained trial acquisition time: 1.69 s ± SD 0.16 s; 95\% CI of mean difference: [0.82, 0.87] s) (Figure \ref{fig:supp 1 - behavioral data}c,d). As designed, the durations of ballistic and sustained trials were largely separable.

Furthermore, the main task assumes that the explanatory variable is the duration of movements, but it does not decouple duration from the cursor gain. To determine whether the result is correlated with duration or with gain, we experimented with a distance-scaling BMI task (see Methods) involving near and far reaches with a unified cursor gain (Figure \ref{fig1:brains tasks & behavioral data}c). We also refer to this task as the radial-grid task. The two types of reaches correspond to two eight-target sets, each with a different distance from the center cue. The participant acquired the targets consistently (near trial acquisition time: 0.75 s ± SD 0.15 s; far trial acquisition time: 1.55 s ± SD 0.40 s; 95\% CI of mean difference: [0.76, 0.86] s) (Figure \ref{fig:supp 1 - behavioral data}e,f). 

Similar to the non-interleaved gain-scaling task, the interleaved gain-scaling and distance-scaling tasks show multiphasic neural dynamics seen in cross-validated neural speed (Figure \ref{fig:supp 2 - neural speeds and peak statistics}c-f) and crossnobis distance (Figure \ref{fig:supp 2 - neural speeds and peak statistics}i-l) with an initial peak during movement onset and subsequent return to baseline. The peak in neural speed share similar behaviors where peak onset time and peak time in ballistic or near trials slightly precede those in sustained or far trials (Figure \ref{fig:supp 2 - neural speeds and peak statistics}m,n) while peak magnitude and duration differ by task condition and brain area (Figure \ref{fig:supp 2 - neural speeds and peak statistics}o,p).

Mid-movement cross-condition discrepancy is observed in interleaved gain-scaling and distance-scaling tasks both in single neuron (Figure \ref{fig2:single and population neurons}e; Figure \ref{fig:supp 3 - perc neurons discrepancy}d-g) and population analyses (Figure \ref{fig3:LDS and population neuron discrepancy}g) as well as inference results (Figure \ref{fig:supp 4 - LDS cross speed inference}c-f). Condition-specific dwell times in the steady phase were observed in time-colored LDS trajectories (Figure \ref{fig:supp 5 - LDS time gradient trajectories}o-z) as well as discrete-state colored rSLDS trajectories and flow fields (Figure \ref{fig:supp 6 - rSLDS discrete state trajectories}o-z). 

\section{Discussion}\label{discussion}
\subsection{Stable dynamics versus flexible control of motor cortex activity}

The motor cortex’s precise mechanisms have been debated for decades. Recent studies have highlighted how the motor cortex (MC) can be modeled as an autonomous pattern generator during arm movement \citep{Churchland2012-hp, Shenoy2013-dn, Pandarinath2015-im, Pandarinath2018-of, Pandarinath2018-kr, Sussillo2015-cz, Vyas2020-an}. Given the similarities between able-bodied movements and brain-machine interface (BMI) control \citep{Hochberg2006-qn, Hwang2013-lu, Golub2016-xg, Guan2022-qw}, an open question was whether autonomous dynamics would also generalize to MC activity during BMI control. 

We found that the two participants generated MC activity during BMI control beyond the constraints of the autonomous dynamical systems hypothesis. During able-bodied-like ballistic movements, MC latent trajectories were similar to previously described oscillatory dynamics (Figure \ref{fig3:LDS and population neuron discrepancy}a,c; Figure \ref{fig:supp 5 - LDS time gradient trajectories}) \citep{Churchland2012-hp}. When necessary for sustained BMI movements, the participant could easily sustain MC activity (Figure \ref{fig3:LDS and population neuron discrepancy}b,d; Figure \ref{fig:supp 5 - LDS time gradient trajectories}) rather than strictly following autonomous dynamics. Interestingly, the MC activity was shared between movements of different conditions, despite the aDSH hypothesis that different initial conditions would lead to different movements \citep{Shenoy2013-dn, Vyas2020-an}. 

Our results indicate that an autonomous dynamical system is too strict a model for the wide temporal scales of possible BMI movements. A similar interpretation was previously reported, although without data \citep{stavisky2016}. 

These results should not be construed to mean that MC has absolutely no constraints on neural activity. A series of studies have examined how the covariance structure of MC during natural movement (also known as a “neural manifold”) constrains one’s ability to generate arbitrary activity \citep{Sadtler2014-sh, Golub2018-fr, Oby2019-tl}. Neural activity patterns within this natural covariance structure are much easier to generate than outside it. However, our results suggest that there are no constraints on the dynamical trajectories to reach the different points on this manifold. 

Because BMI control decouples MC activity from the limb’s biomechanics and natural proprioceptive loop, it enables the question of whether the autonomous dynamics seen in ballistic reaches persist when the temporal demands change. In other words, could the dynamics observed during ballistic reaches simply reflect the behavioral task and its timescale, rather than an intrinsic property of the motor cortex? What may seem like intrinsic dynamics may appear more like behavioral dynamics at different timescales. As shown here, dynamics learned from one behavioral timescale may not apply to similar behaviors performed at shorter or longer timescales. This was also one of the original pitfalls of standard representational modeling (RM); RM found neurons tuned to as many variables as were tested because oftentimes variables are highly correlated. Similarly, without joint behavioral-neural dynamics modeling \citep{Sani2021-xn} or without higher-dimensional tasks \citep{Gao2017-hu}, it can be difficult to disentangle intrinsic neural dynamics from behavioral dynamics.

\subsection{Switching decoders for brain-machine interfaces}

Most previous demonstrations of neural prosthetics have used stationary algorithms like linear regression to decode movement velocity \citep{Collinger2013-ws}. Our results both explain the success of stationary algorithms, like linear regression, as well as outline paths forward for better decoding. Due to the difficulty of controlling modern BMIs, BMI trajectory control often spans multiple seconds, in which sustained neural activity takes precedence over shorter dynamical transients. In this regime, real-time sensitivity \citep{Shanechi2016-xi, Shanechi2017-ms} and fast error correction \citep{Even-Chen2017-yu} are more important than perfect first-time decoding. As BMIs begin to better approximate able-bodied precision \citep{Sussillo2016-ug, Willett2021-vv, Willsey2022-mk}, the sustained component of neural activity becomes shorter, and the dynamical transients take up larger proportions of the movement duration. 


To further improve decoding, decoders may want to take temporal structure into account, for example, explicitly modeling the different phases of BMI movement as motor intent switches from a resting state to a moving state \citep{Sachs2016-ic, Dekleva2021-lr}. Our results suggest that decoders should be sensitive to a minimum of two temporal contexts during movement execution: movement onset (transient phase) and sustained intent (steady phase). An additional movement hold state \citep{Sachs2016-ic, Inoue2018-de} or rest state \citep{Velliste2014-mv} may also be useful for certain applications.

\subsection{Sensory and non-sensory inputs to motor cortex}

Under a dynamical systems framework, the divergence in dynamics between sustained and ballistic movements indicates that the recorded population receives different inputs between the two movement conditions. An open question is where these inputs originate from. During able-bodied reaches, the motor cortex receives rich proprioceptive inputs from S1, which can exhibit similar rotational dynamics \citep{Kalidindi2021-qz}, as well as transformed visual information \citep{Eisenberg2011-ys} potentially from the posterior parietal cortex (PPC). 

Other recent studies have also challenged the autonomous dynamical systems hypothesis by showing that somatosensory input can drive much of the motor cortical activity observed during movement execution \citep{Suresh2020-qp, Kalidindi2021-qz}. For BMI users moving a cursor with a paralyzed effector, as in our study, somatosensory feedback is severed. Are the observed inputs to the motor cortex during sustained activity the direct result of visual feedback or something else? A BMI perturbation study found that visuomotor feedback is initially isolated from the BMI in an output-orthogonal dimension \citep{Stavisky2017-ud}, suggesting that such inputs would occupy a different subspace until the movement goal was updated otherwise. Furthermore, visual feedback of cursor position only weakly affects MC activity in the absence of movement \citep{Stavisky2018-yk}. However, a more targeted study would be necessary to rule out visual feedback as the direct modulation of MC activity. 



\subsection{Unifying neural dynamics and flexible feedback}

Feedback is a core part of flexible movement generation, but the aDSH largely disregards feedback (in the form of external inputs). A complementary framework, optimal feedback control (OFC) \citep{Todorov2002-fm, Scott2004-ju, Shadmehr2008-jn, Franklin2011-om}, has highlighted the importance of feedback for flexible movement generation. Whereas aDSH asserts that movement execution is predetermined by the initial state and intrinsic dynamics, OFC computes movement controls on the fly, taking into account the difference between the current and desired state of effectors. This helps us understand behavior for movements that cannot be predicted at preparation time, such as controlling a new effector \citep{Collinger2013-ws, Wodlinger2015-kb}, rapidly adapting to visual target and obstacle modifications \citep{Dimitriou2013-vx, Nashed2014-cg, Stavisky2017-ud} or load perturbations \citep{Nashed2014-cg, Cluff2015-qa}. Many OFC models posit MC as a feedback controller that integrates information from somatosensory, cerebellar, and other areas to achieve the behavioral goal \citep{Scott2016-ic}. OFC approaches often define a cost function that the system is hypothesized to minimize \citep{Diedrichsen2010-rc}. 

OFC and the dynamical systems literature have often stood in opposition to or in isolation from each other. However, given both the prevalence of dynamical system tools and the importance of feedback to BMI control, clearly, both frameworks provide utility to understanding motor control. A recent review has made an interesting effort to unify OFC and neural dynamical systems (NDS) under the idea of dynamical feedback control (DFC) \citep{Versteeg2022-cl}. Additional experiments may help determine whether this framework can better explain observed phenomena and improve BMI decoding for neuroprosthetic applications.  

\section{Methods}\label{methods}
\subsection{Data collection}

\subsubsection{Study participant}

The study Participant JJ is a right-handed, tetraplegic man. Approximately three years before this study, he sustained a spinal cord injury at cervical level C4-C5. He has residual movement in his upper arms, but he is unable to move or feel his hands. 

The study Participant RD is a right-handed, tetraplegic man. Approximately five years before this study, he sustained a spinal cord injury at the cervical level C3-C4. He shows weak residual movements (twitches) of the wrist and thumbs. He is able to feel tingling sensations when he touches something with his hands, but he can not pinpoint the location of the sensation.

Both participants are part of a BMI clinical study (ClinicalTrials.gov Identifier: NCT01958086). They consented to this study after understanding its nature, objectives, and potential risks. All procedures were approved by the Institutional Review Boards of California Institute of Technology, Casa Colina Hospital and Centers for Healthcare, and the University of California, Los Angeles.

\subsubsection{Multielectrode array implant location}

Participant JJ was implanted with two 96-channel NeuroPort Utah electrode arrays (Blackrock Microsystems) (Figure \ref{fig1:brains tasks & behavioral data}a) about 20 months after injury. One array was implanted near the hand knob area of the left motor cortex (MC). A second array was implanted in the superior parietal lobule (SPL) of the left posterior parietal cortex (PPC). 

Participant RD was implanted with four 64-channel NeuroPort Utah electrode arrays (Figure \ref{fig1:brains tasks & behavioral data}b) about 60 months after injury. Two arrays were implanted near the hand knob area of the left MC. The other two arrays were implanted in the SPL and supramarginal gyrus (SMG) of the left PPC. 

More details regarding the methodology for functional localization and implantation can be found in \cite{Aflalo2015-ik}. All arrays were used for online BMI decoding, although our analyses here only describe data from the MC implants.

\subsubsection{Neural data preprocessing}
Using the NeuroPort system (Blackrock Microsystems), neural signals were recorded from the electrode arrays, amplified, and analog bandpass-filtered (0.3 Hz to 7.5 kHz) before being digitized (30 kHz, 250 nV resolution). A digital high-pass filter (250 Hz) was then applied to each electrode. 

Threshold crossings were detected at a threshold of -3.5 x RMS (root-mean-square of an electrode’s voltage time series). Threshold crossings were used as features for Participant JJ's in-session BMI control during the ballistic-sustained center-out task. For Participant RD's in-session BMI control during both the ballistic-sustained center-out task and the radial-grid tasks, we used neural network-extracted features per electrode \citep{Haghi2024-on}. 

For offline analyses, we used k-medoids clustering on each electrode to spike-sort the threshold-crossing waveforms. The first $n \in \{2, 3, 4\}$ principal components were used as input features to k-medoids, where $n$ was selected for each electrode to account for 95\% of waveform variance. The gap criteria \citep{Tibshirani2001-rd} were used to determine the number of waveform clusters for each electrode. 

Trials were filtered based on cursor behaviors, examined using two metrics: path efficiency and distance-to-target slope. Path efficiency is defined as $\frac{\sum_{t=0}^{t=T-1} \|c_{t+1} - c_{t}\|_2 }{\|c_T\|_2}$, where $c_t$ is the position of the cursor at time $t$, and $T$ is the first time point where the cursor overlaps with the target. Path efficiency measures the straightness of the cursor trajectory during a trial. We discarded trials with path efficiency $> 1.5$. On the other hand, the distance-to-target slope reveals if the cursor is heading toward the target. We discarded trials with positive distance-to-target slopes that occur more than 0.05 screen units (where a screen unit of 1 denotes the length of the shorter side of the screen) away from the origin (starting position of the cursor in each trial). 

We processed the spike-sorted neural data into non-overlapping bins of 10 ms, then applied Gaussian smoothing with a standard deviation (std) of 100 ms and a border factor of 10 to account for edge effects. 

When fitting dynamical models (LDS and rSLDS), we trimmed the data to remove visual delay periods at the beginning of each trial. We removed 139 ms from Participant JJ's trials, 290 ms from Participant RD's trials recorded with the lateral MC array, and 275 ms from Participant RD's trials recorded with the medial MC array. This is done before binning the spike-sorted neural data. We also removed time stamps after target acquisition. 

To visualize more detailed information in crossnobis analyses, we processed the spike-sorted neural data into non-overlapping bins of 1 ms, then applied Gaussian smoothing with an std of 30 ms.

\subsection{Experimental setup}

\subsubsection{Recording sessions}
Experiments were conducted in recording sessions at Casa Colina Hospital and Centers for Healthcare. All tasks were performed with Participant JJ seated in his motorized wheelchair with his hands resting on his lap or an adjacent armrest. Participant JJ viewed text cues on a 27-inch LCD monitor that occupied approximately 40 degrees of visual angle. 

Each session consisted of a series of 2-5 minute, uninterrupted “runs” of the task. The participants rested for several minutes between runs as needed. Table \ref{table1:session summary} summarizes all session data used.

\subsubsection{Center-out-and-back calibration task}
The center-out-and-back task (e.g. \cite{Jarosiewicz2013-jt, Kao2015-sq}) was used to calibrate the decoder for each session. First, a computer-controlled the cursor as it moved out to targets and back to the center. The participants simultaneously attempted to move their thumb or index finger as though they were controlling the cursor via an analog game controller stick. Using this calibration data, we trained a decoder to predict cursor velocity from neural activity. 

After open-loop calibration, the participants performed the center-out-and-back task with a relatively low gain. Partial computer assistance was sometimes applied to the cursor trajectories. The purpose of this follow-up task was to collect more data to train the decoder, as recommended by \citep{Jarosiewicz2013-jt}. 

In some sessions, the participants repeated the task with a small number of repetitions so they could familiarize himself with the decoder behavior \citep{Willett2017-kx}.

\subsubsection{Center-out brain-machine interface task with variable gain-scaling}
The participants used cursor BMIs to complete the center-out task \citep{Georgopoulos1982-xf, Jarosiewicz2013-jt} under two decoder gain parameters: high gain for ballistic reaches and low gain for sustained reaches. We refer to this task as the ballistic-sustained center-out task. The reach duration was tuned to complete in approximately 500 ms (ballistic) or 2 seconds (sustained) for Participant JJ using a velocity gain parameter of the neural decoder \citep{Willett2017-kx}. Because Participant RD is not able to consistently perform ballistic reaches with completion times as short as 500 ms, we instead tuned his ballistic reaches to complete in approximately 750 ms. The ballistic-movement condition is designed to resemble prior NHP reaching experiments. Since closed-loop decoders cannot match the speed and accuracy of NHP behavior, we add computer assistance (see below Methods) to allow ballistic movements without the need for online corrections.  

\subsubsection{Interleaved center-out brain-machine interface task with variable gain-scaling}
The interleaved center-out task is identical to the center-out brain-machine interface task with variable gain scaling, except that its trials are interleaved. Only Participant RD attempted this task. The goal of this task is to control for the effect of anticipation since the participant can no longer anticipate the cursor gain of the upcoming trial. We occasionally change the number of trials under different conditions, which the participant did not realize. 

\subsubsection{Near-Far center-out brain-machine interface task with variable distance}
We performed an additional center-out task variant to simulate sustained versus ballistic reaches. We refer to this variant as the radial-grid task. In this task, targets were positioned at two different distances from the center (Figure \ref{fig1:brains tasks & behavioral data}c). The participant was instructed to move the cursor to the target as soon as the target appeared and then relax once the target was reached.  Near targets resulted in shorter movement durations, whereas far targets resulted in longer movement durations. This is functionally similar to modulating the decoder gain directly \citep{Willett2017-kx}, as was done in the previous ballistic-sustained task. Only Participant RD attempted this task. This task was only tested with interleaved trials.

\subsubsection{Computer-assisted control: weighted average}
Weighted-average computer assistance blends the decoded velocity with a vector directly toward the target. This type of computer assistance was previously described as “attraction assistance” \citep{Velliste2008-pw}. Assistance values ranged from 0 (full BMI control) to 0.5 (assisted BMI control), where a value of 1 corresponds to full computer control. 

This type of computer-assisted control was used by Participant RD during both the gain-scaling ballistic-sustained center-out task and the variable distance radial-grid task.

\subsubsection{Computer-assisted control: error rail}
Error rail attenuates the components of the decoded velocity that are not directly toward the target. (We denote the direct cursor-to-target vector as $\vec{d}$). This is accomplished by decomposing the decoding cursor velocity $\vec{v}$ into the component parallel to the target direction $\vec{v}_\parallel$ and the component orthogonal to the target direction $\vec{v}_\perp$. With assistance value K, the assisted cursor velocity $\hat{v}$ is  

\begin{align*}
    \hat{v}_\perp &= (1 - K) \vec{v}_\perp \\
    \hat{v}_\parallel &= 
    \begin{cases} 
        \vec{v}_\parallel & \text{if } \vec{v}_\parallel \cdot \vec{d} > 0 \\
        (1 - K) \vec{v}_\parallel & \text{otherwise}
    \end{cases} \\
    \hat{v} &= \hat{v}_\perp + \hat{v}_\parallel
\end{align*}

The assistance value $K$ was specified for each run and ranged from 0 (full BMI control) to 0.3, where a value of 1 restricts the decoder to motion directly towards the target. Error rail is similar in some ways to ortho-impedance \citep{Collinger2013-ws, Wodlinger2015-kb} but also attenuates parallel-component velocities in the reverse direction. 

This type of computer-assisted control was used by Participant JJ.

\subsection{Closed-loop decoding pipelines}

\subsubsection{Neural dynamical filter}
During Participant JJ's center-out sessions, we preprocessed the neural activity by binning spike counts at non-overlapping 30-ms bins, z-scoring the firing rates for each channel, and reducing the dimensionality to the first 15 principal components. We decoded movement intent from the reduced-dimensionality population activity using the neural dynamical filter (NDF) \citep{Kao2015-sq} with a 10-dimensional latent state. NDF learns a latent-state linear dynamical system of the neural population activity. For online decoding, NDF linearly predicts kinematics from the dynamics-filtered latent states. The NDF is described in detail in \cite{Kao2015-sq}. 

\subsubsection{Linear decoder with deep-learning-based feature extraction}
In later tasks, recordings yielded relatively few high-SNR waveforms, so we switched from threshold-crossing rates to broadband features. We used a temporal convolutional neural network (denoted ``FENet") to extract features from 30kHz-sampled raw voltage time series \citep{Haghi2024-on}. 


Before inputting the FENet features into the linear decoder, we preprocessed them in a series of steps. First, we z-scored input features. Next, to prevent unexpected channel noise from disproportionately degrading decoding, we bound the z-scored values between [-3, 6] (s). Because FENet generates multiple features (K=8) per electrode, we used partial least squares regression to reduce this number to two informative features (K=2) for each electrode. Next, we reduced feature dimensionality by using partial least squares to predict the input features, which were smoothed using an 800-ms minimum-jerk kernel; this approach is analogous to a linear autoencoder and helps denoise data that is expected to be autocorrelated. Finally, the firing rates were smoothed with an exponential filter (time-constant = 585ms)

\subsection{Statistical analysis}

\subsubsection{Cross-validated neural speed and crossnobis distance}
\cite{Kao2015-sq} defines neural population speed as $\|r_{k+1} - r_k\|_2$. We use a similar definition here but address one statistical drawback in the original formulation: noise biases Euclidean distances upward \citep{Walther2016-gn}. In other words, adding independent Gaussian noise to the firing rates increases $\|r_{k+1} - r_k\|_2$, even though the neural population speed doesn’t appear to change otherwise. To find an unbiased estimate of the neural speed, it is useful to cross-validate speed estimates across independent partitions of trial repetitions \citep{Walther2016-gn}:

\begin{equation}
    \|\dot{r}\|^2_{cv} = \left[ \frac{r_{k+1} - r_k}{\Delta t} \right]_A \left( \frac{\Sigma_A + \Sigma_B}{2} \right)^{-1} \left[\frac{r_{k+1} - r_k}{\Delta t} \right]_B^T
\end{equation}

where $A$ and $B$ indicate independent partitions of the trials, $\Sigma$ is the (regularized \citep{Ledoit2003-yy} noise covariance matrix, $r_k$ are the firing rate vectors for time index $k$ stacked across the respective partition’s trials, and $\Delta t$ is the time difference between consecutive time indices. The units of $\|\dot{r}\|^2_{cv}$ are $(Hz / s)^2$. Although one can take the signed square root to obtain more meaningful units, here we prioritize the unbiased properties of the metric \citep{Schutt2019rsatoolbox}. 

The cross-validated neural population speed $\|\dot{r}\|^2_{cv}$ is a modified version of the cross-validated Mahalanobis distance (or crossnobis distance) \citep{Walther2016-gn}. It estimates the noise-normalized magnitude of the neural velocity that is consistent across trial repetitions. Here, we often abbreviate this metric as “neural speed”.

We also use the unmodified crossnobis distance without the concept of time to analyze the differences in tuning across phases of a trial: 

\begin{equation}
    \|\Delta r\|^2_{cv} = \left[ r_{k_1} - r_{k_2} \right]_A \left( \frac{\Sigma_A + \Sigma_B}{2} \right)^{-1} \left[r_{k_1} - r_{k_2} \right]_B^T
\end{equation}

where $r_{k_1}$ and $r_{k_2}$ are firing rate vectors at two time indices that are not necessarily consecutive. The units of $\|\Delta r\|^2_{cv}$ are $Hz^2$. 

In visualizations of cross-validated neural speed (Figure \ref{fig2:single and population neurons}a; Figure \ref{fig:supp 2 - neural speeds and peak statistics}a-f) and crossnobis distance (Figure \ref{fig2:single and population neurons}b; Figure \ref{fig:supp 2 - neural speeds and peak statistics}g-l), we preprocessed spike-sorted data by applying Gaussian smoothing with an std of 30 ms and border factor of 10 (to account for edge effects).  

We define the peak time as the time (after trial start) at which the initial peak of cross-validated neural speed occurs. We use the full width at half maximum to define the duration of this peak. We define the onset time of the initial peak as the last timestamp where the cross-validated neural speed exceeds 1.5 times the baseline standard deviation in each trial. The baseline is defined as the last 200 ms of the inter-trial interval (ITI).

\subsubsection{Visual delay period}
We refer to the goal-directed motor responses to visual signals defined in \cite{Scott2016-ic} as the visual delay period. We compute the visual delay period as the time interval between the trial start and the peak onset in cross-validated neural speed.

\subsection{Dynamical models}
We applied recurrent switching linear dynamical systems (rSLDS) \citep{pmlr-v54-linderman17a} on data collected from both participants. Specifically, we applied the SLDS class with recurrent transitions, Gaussian dynamics, single-subspace orthogonal Gaussian emissions, and autoregressive hidden Markov model (ARHMM) initialization.

We swept the number of continuous states (C $\in \{$2, 3, ..., 20$\}$) and discrete states (D $\in \{$1, 2, 3, 4, 6, 8, 12, 16$\}$) with five-fold cross-validation. When $d = 1$, the model becomes equivalent to a linear dynamical system (LDS). To evaluate the test performance, we used four metrics (Figure \ref{fig:supp 8 - numerical results}): 1) absolute angle error of decoded position error \citep{Willett2018-oc}, 2) dynamical state update process (DSUP) ratios \citep{Kao2015-sq}, 3) variance explained in forecast inference, and 4) dimensional entropy difference.

\subsubsection{Position error}
The position error is an intention estimation method. Since the participants move the cursor to one of eight targets in each trial, a ground truth direction exists for the movement from each cursor position. We can decode the direction at each time point and compare its absolute difference to the ground truth.

\subsubsection{Dynamical state update process (DSUP) ratio}
The DSUP ratio \citep{Kao2015-sq} measures the proportion of state updates explained by the dynamical model as:

\begin{equation}
    r_t = \frac{\|(A_{z_{t-1}} - I) \hat{x}_{t-1} + b_{z_{t-1}}\|}{\|(A_{z_{t-1}} - I) \hat{x}_{t-1} + b_{z_{t-1}}\| + \|K_t (y_t - C_{z_t}(A_{z_{t-1}} \hat{x}_{t-1} + b_{z_{t-1}}) -  d_{z_t})\|}
    \label{eqt:DSUPR}
\end{equation}

where $\|(A_{z_{t-1}} - I) \hat{x}_{t-1} + b_{z_{t-1}}\|$ is the contribution by the dynamical model and the other quantity in the denominator is called the innovations process which models what cannot be explained in the neural data. Also, $K_t$ and $\hat{x}_t$ are the Kalman gain and Kalman filter estimates of the continuous state, respectively. $\Sigma_t$, $N$, and $R$ are the covariance matrices of the continuous states, dynamics, and observations, respectively.

\subsubsection{Forecast inference}
Forecast inference predicts the next observation $\hat{y}_{t+1}$ from the current continuous state $x_t$ by applying the learned dynamics and emission parameters as: 

\begin{align}
    \hat{x}_{t+1} &= A_{z_t} x_t + b_{z_t} \\
    \hat{y}_{t+1} &= C_{z_{t+1}} \hat{x}_{t+1} + d_{z_{t+1}}
    \label{eq:forecast_inference}
\end{align}

We can then compute the R2 difference between the predicted observation and the data.

\subsubsection{Dimensional entropy difference}
We devised the dimensional entropy difference metric to quantify whether the discrete states learned by rSLDS consistently switch from one to another across trials. The metric favors having a single dominant state at each time point (i.e., low time-wise entropy) and penalizes the case where a single state dominates an entire trial (i.e., low trial-wise entropy). In other words, the metric rewards the scenario where: 1) at each time point $t$, most trials use the same state (yielding low entropy across trials at that time), and 2) within each trial $i$, the trial visits multiple states (yielding a high entropy over time within that trial). Mathematically, this means minimizing the average time-wise entropy while maximizing the average trial-wise entropy:

\begin{align}
    H_i &= -\frac{1}{\ln{n_z}} \sum_{z=0}^{n_z} p_z(i) \ln(p_z(I)) \\
    H_t &= -\frac{1}{\ln{2}} [\max_z p_z(t) \ln{(\max_z p_z(t))} + (1 - \max_z p_z(t)) \ln{(1 - \max_z p_z(t))}] \\
    H &= \frac{1}{n_{trials}} \sum_i H_i - \frac{1}{n_{times}} \sum_t H_t 
    \label{eqt:dimensional_entropy_difference}
\end{align}

where $p_z (i)$ is the fraction of time steps in trial $i$ spent in state $z$, and $p_z (t)$ is the fraction of trials in which state $z$ appears at time $t$. Because each entropy term ($H_i$ and $H_t$) is normalized to $[0, 1]$, the dimensional entropy difference $H$ falls in $[-1, 1]$.

\subsubsection{Model selection}
Overall, position error and forecast inference (Figure \ref{fig:supp 8 - numerical results}) favor larger numbers of continuous states while being insensitive to discrete states. For DSUP ratios, having more discrete states is beneficial, but the performance gain diminishes as the number of discrete states increases, with the gain from D1 (LDS) to D2 being the most dramatic. For dimensional entropy difference, having two discrete states is significantly more favorable. We chose C3D2 for our analyses as this combination is simplistic, unlikely to overfit, facilitates visualization, and performs reasonably well in the metrics. Compared to LDS (D1), rSLDS (D2) performs similarly on position error as well as forecast and better on DSUP ratios. Compared to rSLDS (D $\geq 3$), rSLDS (D2) performs significantly better on dimensional entropy difference.

\section*{Acknowledgments}
We thank participant JJ and participant RD for making this research possible. We also thank Viktor Scherbatyuk for technical assistance. 

This work was supported by NIH National Eye Institute Awards UG1EY032039, the T\&C Chen Brain-Machine Interface Center, the James G. Boswell Foundation, and the Swartz Foundation.

\section*{Author contributions}
F.Y., C.G., and T.A. conceived the project. F.Y., C.G., and T.A. designed the BMI experiments. F.Y., C.G., and J.G. collected neural data during experimental sessions. F.Y. and C.G. wrote data processing code. F.Y. wrote custom analysis and visualization code with guidance from C.G. and J.G. F.Y. and C.G. drafted the manuscript; all authors discussed results and edited the final version. J.G. and K.P. coordinated the scheduling and logistics of participants. K.P. managed institutional approvals and coordinated clinical logistics. E.R. established and managed the hospital study site. C.L. and A.B. provided long-term medical supervision for the participants. A.B. conducted neurosurgery to implant recording electrode arrays. R.A. obtained funding and provided overall supervision.

\section*{Competing interests}
The authors declare no competing interests.

\bibliography{sn-bibliography}

\clearpage
\appendix

\renewcommand{\thefigure}{S\arabic{figure}}
\renewcommand{\thetable}{S\arabic{table}}
\setcounter{figure}{0}
\setcounter{table}{0}

\renewcommand{\theHfigure}{S\arabic{figure}}
\renewcommand{\theHtable}{S\arabic{table}}

\section*{Supplementary Figures and Tables}

\begin{sidewaystable}[ht!]
\caption{Summary of data used across all sessions.}\label{table1:session summary}
\begin{tabular}{cccccccc}
\toprule
\textbf{Session Date} & \textbf{Participant} & \textbf{Task}         & \textbf{Array Location} & \textbf{\# Neurons} & \textbf{Conditions}  & \textbf{\# Trials 1} & \textbf{\# Trials 2} \\
\midrule
20190412              & JJ                   & center-out            & MC                      & 134                 & ballistic, sustained & 58                             & 58                             \\
20190517              & JJ                   & center-out            & MC                      & 127                 & ballistic, sustained & 58                             & 71                             \\
20190528              & JJ                   & center-out            & MC                      & 146                 & ballistic, sustained & 64                             & 78                             \\
\multicolumn{4}{c}{\textbf{JJ SUM}}                                                            & \textbf{407}        & \textbf{/}           & \textbf{180}                   & \textbf{207}                   \\
20240516              & RD                   & center-out            & MC-LAT                  & 63                  & ballistic, sustained & 55                             & 64                             \\
20240516              & RD                   & center-out            & MC-MED                  & 78                  & ballistic, sustained & 55                             & 64                             \\
20240530              & RD                   & center-out            & MC-LAT                  & 88                  & ballistic, sustained & 50                             & 64                             \\
20240530              & RD                   & center-out            & MC-MED                  & 43                  & ballistic, sustained & 50                             & 64                             \\
20240816              & RD                   & center-out            & MC-LAT                  & 90                  & ballistic, sustained & 112                            & 96                             \\
20240816              & RD                   & center-out            & MC-MED                  & 79                  & ballistic, sustained & 112                            & 96                             \\
20240820              & RD                   & center-out            & MC-LAT                  & 85                  & ballistic, sustained & 64                             & 64                             \\
20240820              & RD                   & center-out            & MC-MED                  & 82                  & ballistic, sustained & 64                             & 64                             \\
20241015              & RD                   & center-out            & MC-LAT                  & 88                  & ballistic, sustained & 103                            & 96                             \\
20241015              & RD                   & center-out            & MC-MED                  & 101                 & ballistic, sustained & 103                            & 96                             \\
20241022              & RD                   & center-out            & MC-LAT                  & 91                  & ballistic, sustained & 89                             & 96                             \\
20241022              & RD                   & center-out            & MC-MED                  & 75                  & ballistic, sustained & 89                             & 96                             \\
20241105              & RD                   & radial-grid           & MC-LAT                  & 103                 & near, far            & 93                             & 92                             \\
20241105              & RD                   & radial-grid           & MC-MED                  & 51                  & near, far            & 93                             & 92                             \\
20241211              & RD                   & radial-grid           & MC-LAT                  & 91                  & near, far            & 76                             & 78                             \\
20241211              & RD                   & radial-grid           & MC-MED                  & 52                  & near, far            & 76                             & 78                             \\
20250408              & RD                   & radial-grid           & MC-LAT                  & 73                  & near, far            & 113                            & 120                            \\
20250408              & RD                   & radial-grid           & MC-MED                  & 45                  & near, far            & 113                            & 120                            \\
20250417              & RD                   & center-out-interleave & MC-LAT                  & 71                  & ballistic, sustained & 189                            & 152                            \\
20250417              & RD                   & center-out-interleave & MC-MED                  & 70                  & ballistic, sustained & 189                            & 152                            \\
20250422              & RD                   & center-out-interleave & MC-LAT                  & 61                  & ballistic, sustained & 175                            & 112                            \\
20250422              & RD                   & center-out-interleave & MC-MED                  & 86                  & ballistic, sustained & 175                            & 112                            \\
20250509              & RD                   & center-out-interleave & MC-LAT                  & 67                  & ballistic, sustained & 164                            & 144                            \\
20250509              & RD                   & center-out-interleave & MC-MED                  & 76                  & ballistic, sustained & 164                            & 144                            \\
\multicolumn{4}{c}{\textbf{RD MC-LAT SUM}}                                                     & \textbf{971}        & \textbf{/}           & \textbf{1283}                  & \textbf{1178}                  \\
\multicolumn{4}{c}{\textbf{RD MC-MED SUM}}                                                     & \textbf{838}        & \textbf{/}           & \textbf{1283}                  & \textbf{1178}                  \\
\botrule
\end{tabular}
\end{sidewaystable}

\begin{figure}[ht!]
\centering
\includegraphics[width=\linewidth]{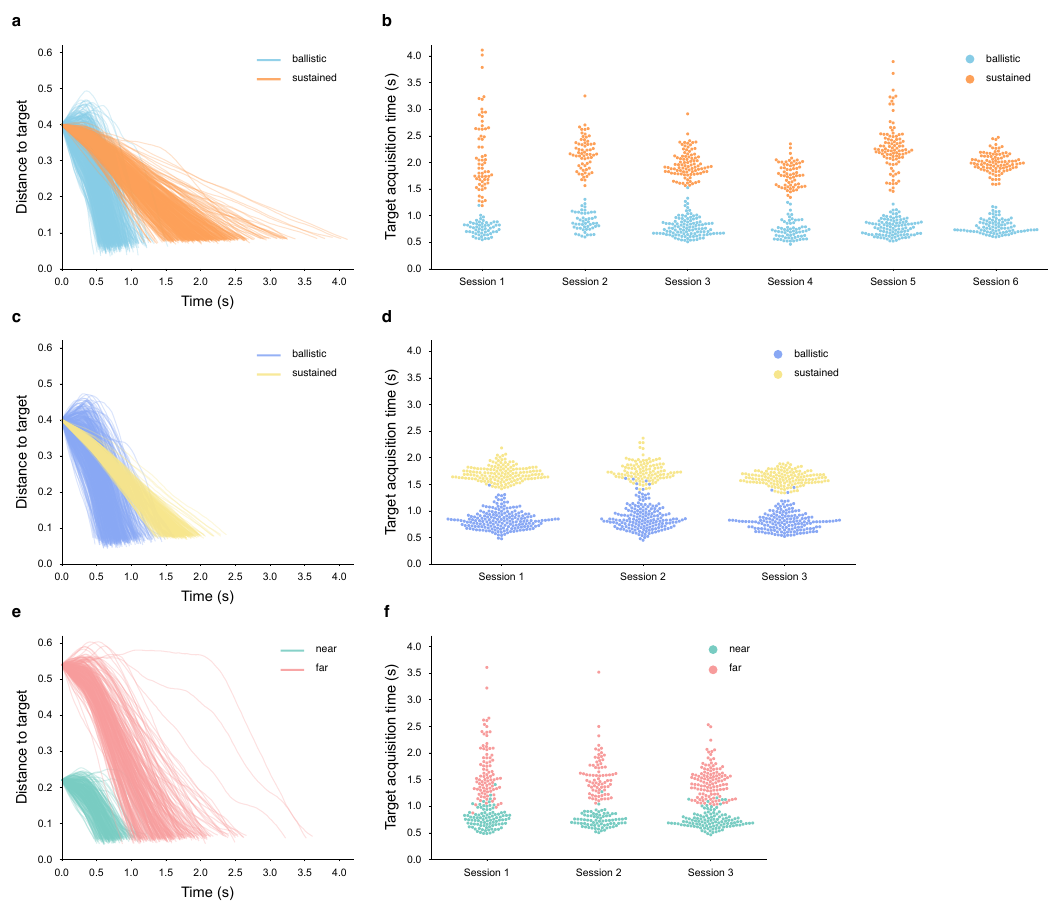}
\caption{\textbf{a,c,e}, Distance to target over time for Participant RD’s center-out task (\textbf{a}), interleaved center-out task (\textbf{c}), and radial-grid task (\textbf{e}) trials. Each trace represents a trial. \textbf{b,d,f}, Target acquisition times for Participant RD’s center-out task (\textbf{b}), interleaved center-out task (\textbf{d}), and radial-grid task (\textbf{f}) trials.
Each marker represents a trial.}
\label{fig:supp 1 - behavioral data}
\end{figure}

\begin{figure}[ht!]
\centering
\includegraphics[width=\linewidth]{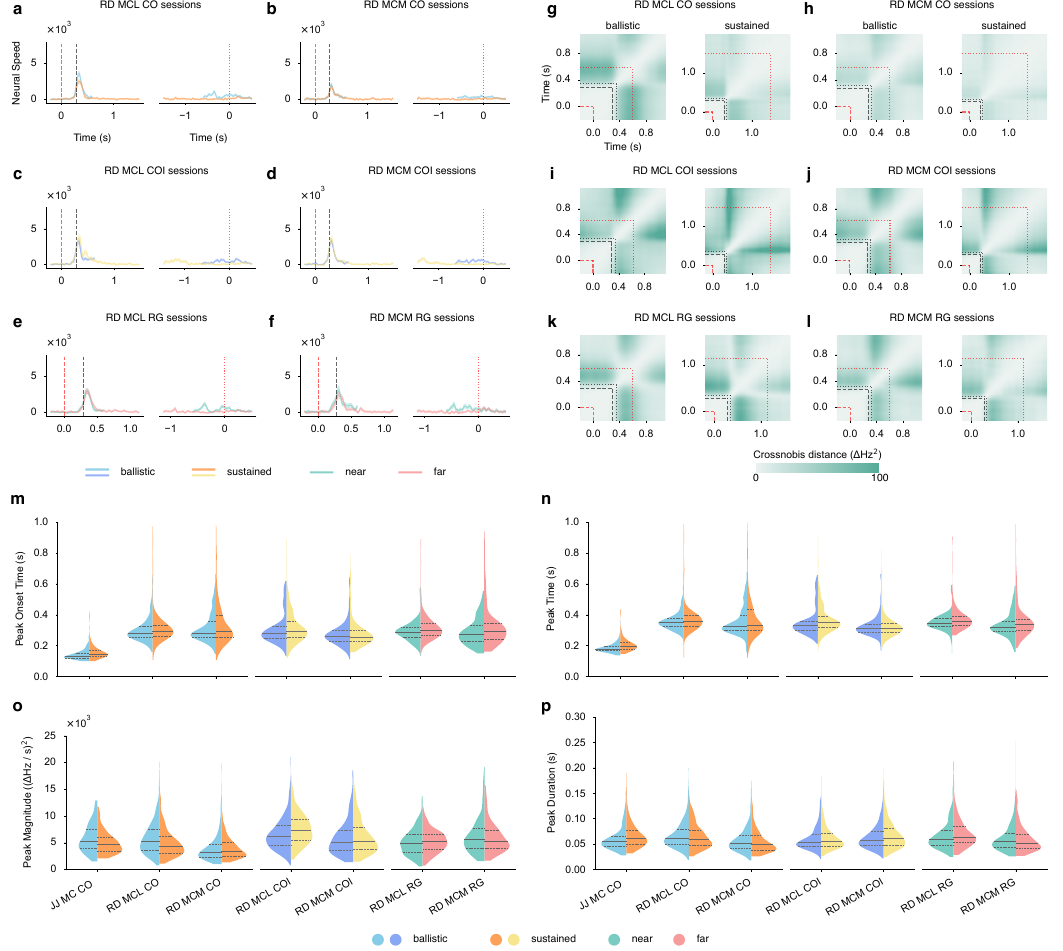}
\caption{\textbf{a-f}, Cross-validated neural speed during trials under different conditions in Participant RD’s center-out task (\textbf{a,b}), interleaved center-out task (\textbf{c,d}), and radial-grid task (\textbf{e,f}) sessions in the same format as Figure 2a. Trials of each condition are truncated to the shortest trial duration observed for that condition. The error bands show 95\% confidence intervals across trials. Left (in each panel): Trials are aligned to the start time marked by the red dashed line. The black dashed line (left) marks the peak onset time (see Methods). Data includes 200 ms before trial start. Right (in each panel): Trials are aligned to the target acquisition time marked by the red dotted line. Data includes 500 ms after target acquisition. \textbf{g-l}, Crossnobis distance across time during trials under different conditions in Participant RD’s center-out task (\textbf{g,h}), interleaved center-out task (\textbf{i,j}), and radial-grid task (\textbf{k,l}) sessions in the same format as Figure 2b. Data includes 200 ms before trial start and 500 ms after target acquisition. Trials of each condition are truncated to the shortest trial duration observed for that condition. The red dashed lines mark the trial start time. The red dotted lines mark the target acquisition time of the shortest trial of each condition. The black dashed lines and dotted lines mark the peak onset times and peak times (see Methods), respectively, in a of each condition. \textbf{m-p}, Peak onset time (\textbf{m}), peak time (\textbf{n}), peak magnitude (\textbf{o}), and peak duration (\textbf{p}) of cross-validated neural speed for all sessions, participants, brain areas, and task conditions.}
\label{fig:supp 2 - neural speeds and peak statistics}
\end{figure}

\begin{figure}[ht!]
\centering
\includegraphics[width=\linewidth]{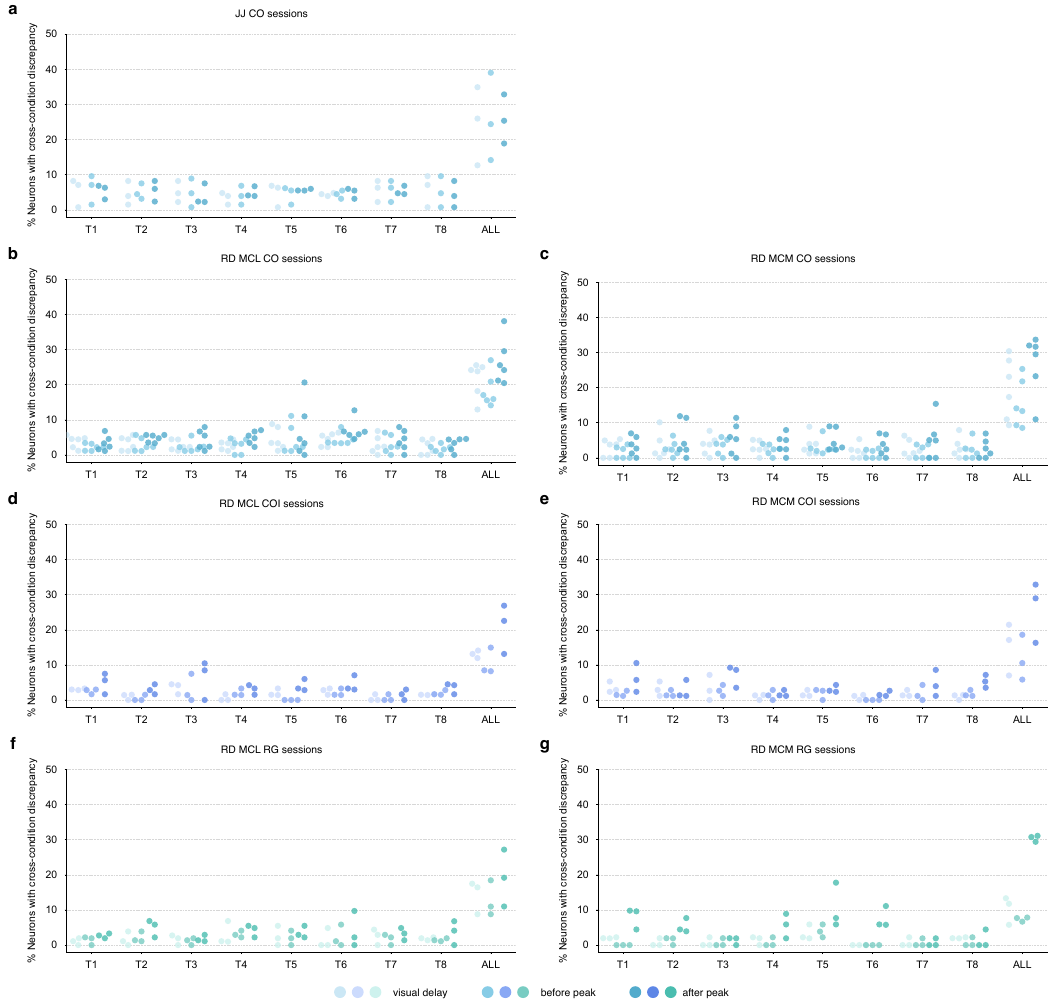}
\caption{Percent of neurons with significant cross-condition discrepancy across targets (T1 to T8) in all sessions from both participants. The ALL columns show unions across targets. Figure 2e summarizes results from all targets and disregards the ALL columns. Blues, indigos, and greens denote the center-out, interleaved center-out, and radial-grid tasks, respectively. Lighter to darker shades of each color correspond to the visual delay (see Methods), before-peak, and after-peak windows, respectively. \textbf{a}, Results from Participant JJ's center-out sessions. \textbf{b,c}, Results from Participant RD's center-out sessions. \textbf{d,e}, Results from Participant RD's interleaved center-out sessions. \textbf{f,g}, Results from Participant RD's radial-grid sessions.}
\label{fig:supp 3 - perc neurons discrepancy}
\end{figure}

\begin{figure}[ht!]
\centering
\includegraphics[width=\linewidth]{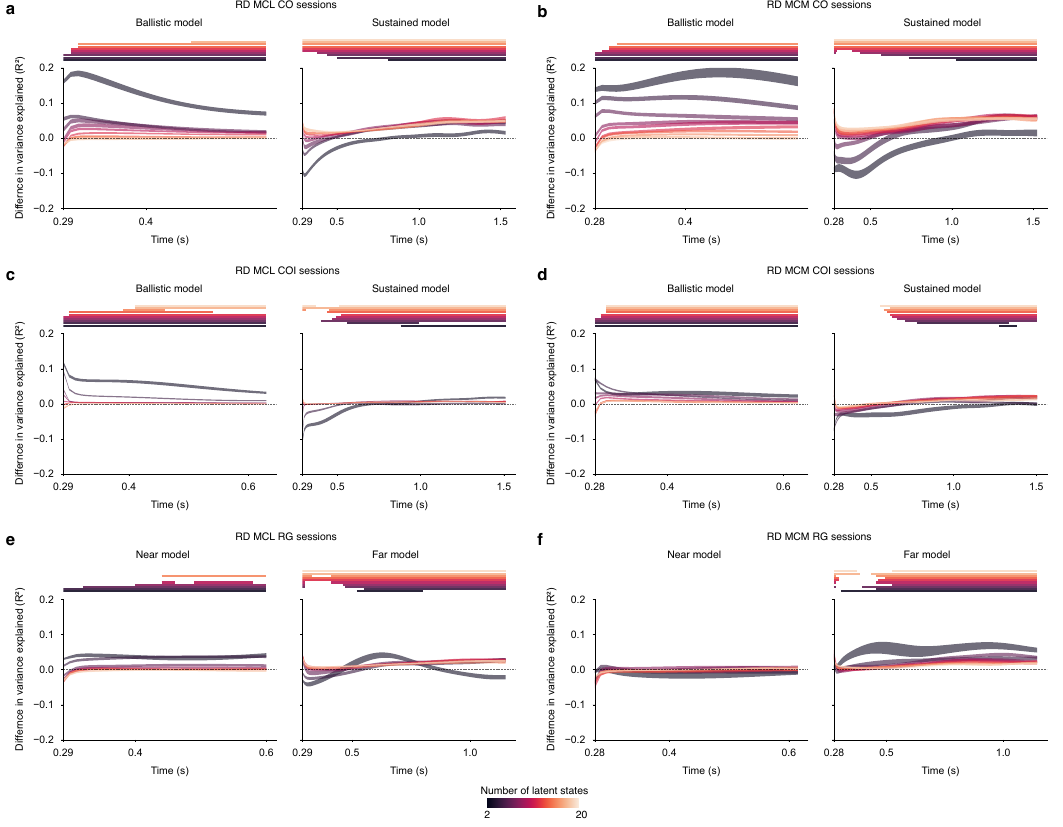}
\caption{SEMs of difference in performance between same-condition and cross-condition inference of trials under different conditions from Participant RD's sessions. Positive values on the y-axis denote better performance achieved by the same-condition inference. Horizontal bars indicate significantly better performance achieved by same-condition inference (Wilcoxon signed-rank test with false discovery rate correction, $P < 0.05$). \textbf{a,b}, Results from Participant RD's center-out sessions. \textbf{c,d}, Results from Participant RD's interleaved center-out sessions. \textbf{e,f}, Results from Participant RD's radial-grid sessions.}
\label{fig:supp 4 - LDS cross speed inference}
\end{figure}

\begin{figure}[ht!]
\centering
\includegraphics[width=\linewidth]{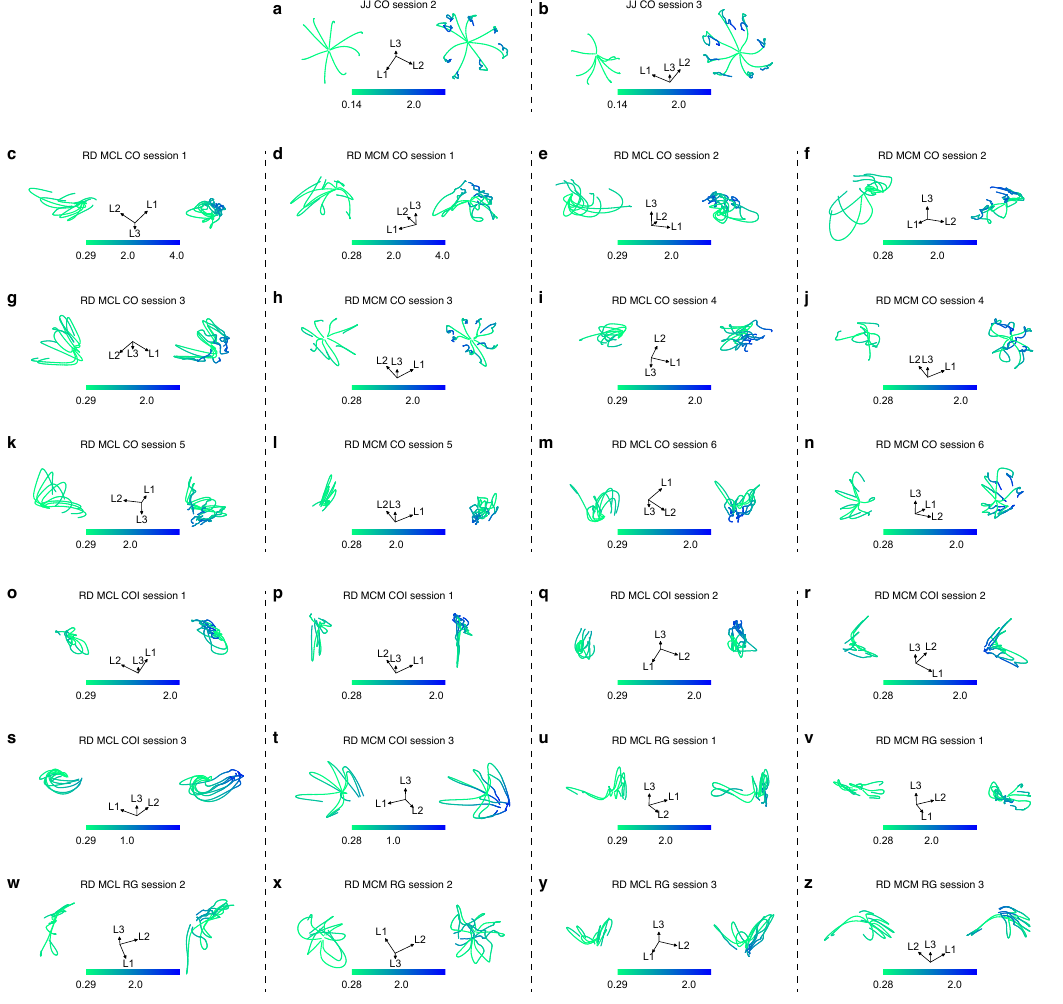}
\caption{LDS latent neural trajectories color-coded by time since the start of trials. The trajectories show target averages after time-resampling. \textbf{a,b}, Results from Participant JJ's center-out sessions. \textbf{c-n}, Results from Participant RD's center-out sessions. \textbf{o-t}, Results from Participant RD's interleaved center-out sessions. \textbf{u-z}, Results from Participant RD's radial-grid sessions.}
\label{fig:supp 5 - LDS time gradient trajectories}
\end{figure}

\begin{figure}[ht!]
\centering
\includegraphics[width=\linewidth]{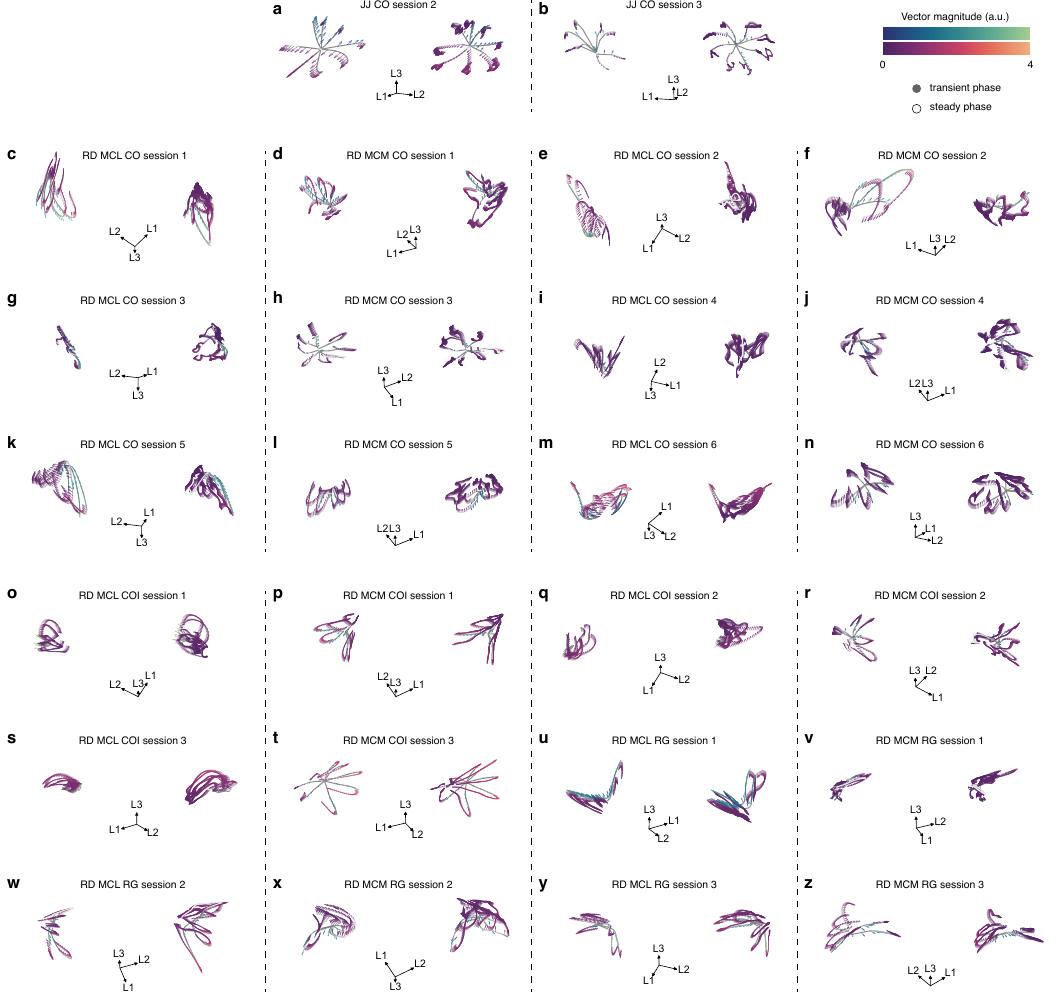}
\caption{rSLDS (C3D2) latent neural trajectories color-coded by discrete states. The trajectories show target averages after time-resampling. Flow field vectors learned by rSLDS are overlaid on the trajectories. Transient phase and steady phase flow field vectors are represented by cold and warm color maps, respectively. \textbf{a,b}, Results from Participant JJ's center-out sessions. \textbf{c-n}, Results from Participant RD's center-out sessions. \textbf{o-t}, Results from Participant RD's interleaved center-out sessions. \textbf{u-z}, Results from Participant RD's radial-grid sessions.}
\label{fig:supp 6 - rSLDS discrete state trajectories}
\end{figure}

\begin{figure}[ht!]
\centering
\includegraphics[width=\linewidth]{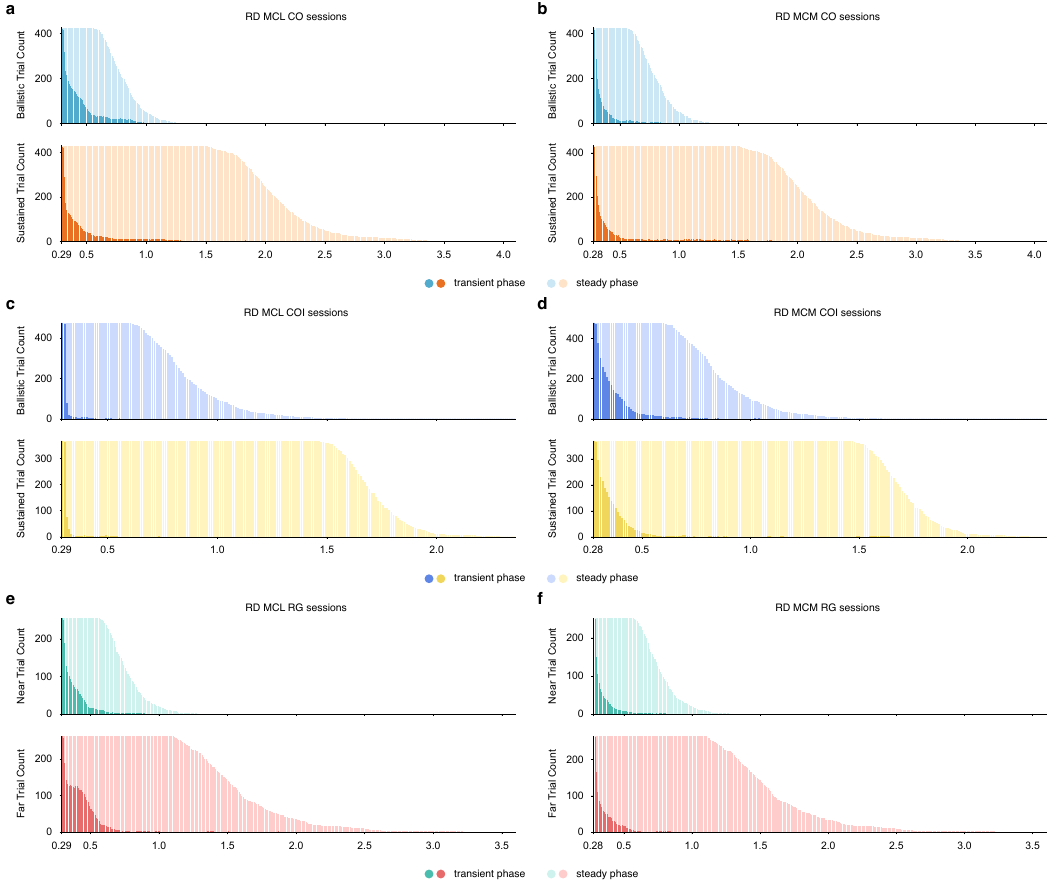}
\caption{Trial counts of rSLDS (C3D2) discrete states over time grouped by task condition. \textbf{a,b}, Results from Participant RD's center-out sessions. \textbf{c,d}, Results from Participant RD's interleaved center-out sessions. \textbf{e,f}, Results from Participant RD's radial-grid sessions.}
\label{fig:supp 7 - rSLDS discrete state over time}
\end{figure}

\begin{figure}[ht!]
\centering
\includegraphics[width=\linewidth]{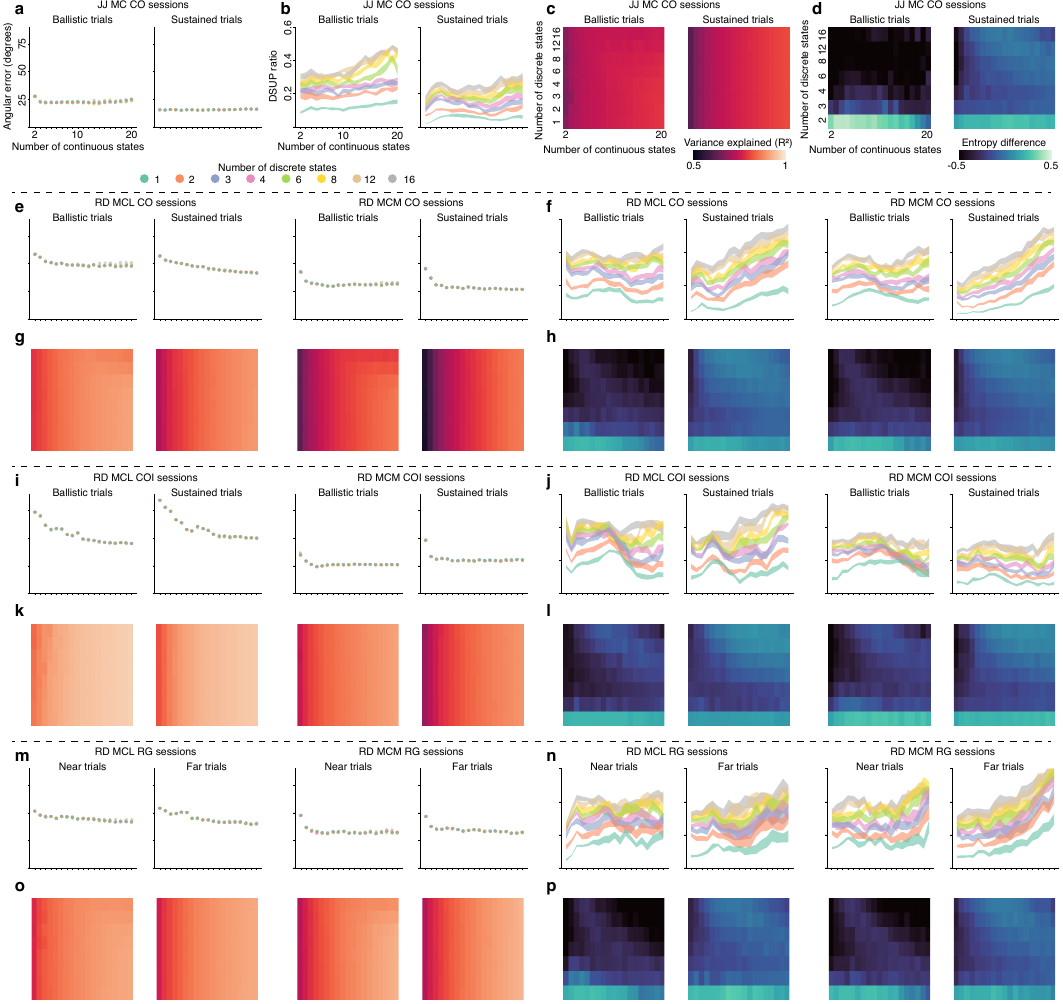}
\caption{Numerical results of LDS and rSLDS on four performance metrics: position error decoding, DSUP ratio, inference, and dimensional entropy difference (see Methods), evaluated over a grid of continuous- and discrete-state counts. Results are organized by participant, brain area, and task. \textbf{a,e,i,m}, Position-error decoding results. \textbf{b,f,j,n}, DSUP ratio results. \textbf{c,g,k,o}, Inference results. \textbf{d,h,l,p}, Dimensional entropy difference results.}
\label{fig:supp 8 - numerical results}
\end{figure}

\end{document}